\documentclass[twocolumn]{revtex4-1}
\usepackage{hyperref}
\usepackage{amssymb} 
\usepackage{amsmath}
\usepackage{amsfonts}
\usepackage{bm}
\usepackage{dcolumn}
\usepackage{graphicx}
\usepackage{subfigure}

\begin{document}

\title{Empirical correction of a toy climate model}
\author{Nicholas A. Allgaier, Kameron D. Harris, and Christopher M. Danforth}
\affiliation{Department of Mathematics and Statistics,  Vermont Advanced Computing Center, Complex Systems Center, The University of Vermont}
\date{\today} 

\begin{abstract}
Improving the accuracy of forecast models for physical systems such as the atmosphere is a crucial ongoing effort. Errors in state estimation for these often highly nonlinear systems has been the primary focus of recent research, but as that error has been successfully diminished, the role of model error in forecast uncertainty has duly increased. The present study is an investigation of a particular empirical correction procedure that is of special interest because it considers the model a ``black box'', and therefore can be applied widely with little modification. The procedure involves the comparison of short model forecasts with a reference ``truth'' system during a training period in order to calculate systematic (1) state-independent model bias and (2) state-dependent error patterns. An estimate of the likelihood of the latter error component is computed from the current state at every timestep of model integration. The effectiveness of this technique is explored in two experiments: (1) a perfect model scenario, in which models have the same structure and dynamics as the true system, differing only in parameter values; and (2) a more realistic scenario, in which models are structurally different (in dynamics, dimension, and parameterization) from the target system. In each case, the results suggest that the correction procedure is more effective for reducing error and prolonging forecast usefulness than parameter tuning. However, the cost of this increase in average forecast accuracy is the creation of substantial qualitative differences between the dynamics of the corrected model and the true system. A method to mitigate the structural damage caused by empirical correction and further increase forecast accuracy is presented. 
\end{abstract}

\maketitle

\section{Introduction}
Advances in computational power and increasingly accurate techniques for estimating the current state of the Earth's atmosphere have significantly improved numerical weather prediction (NWP) \cite{anderson2001,whitaker2002,ott2004}. As state estimation error is reduced due to improved methods of data assimilation, error in the model tendency plays an increasing role in the uncertainty of predictions at every temporal and physical scale \cite{hamill2000,houtekamer2001,kalnaybook2003}. 

In 1978, Leith introduced a statistical technique to correct model tendency error, in which short model forecasts are compared to a time series of reference ``truth'' to estimate both state-independent model bias, and state-dependent error components which are approximated by a least-squares linear function of the model state \cite{leith1978}. More recently, empirical correction has been employed with success in atmospheric models with relatively few degrees of freedom (e.g.\ $N = O(10^2)$ in \cite{delsole1999}), and low-dimensional modifications of the technique involving, for example, singular value decomposition (SVD) of the state-dependent correction operator (the least-squares linear function proposed by Leith) have proven successful in models with as many has $O(10^5)$ degrees of freedom \cite{danforth2007}. In this study, we apply the original technique developed by Leith to simple three-dimensional Lorenz models \cite{lorenz1963}, where in addition to testing the effectiveness of empirical correction, we aim to understand the dynamical ramifications of a statistical approach to the correction of model tendency error.

The model tendency $M(\mathbf{x})$ is defined as the change in state variables over one timestep of numerical integration, which we denote as a time derivative:
\begin{equation}
M(\mathbf{x}) \;\equiv\; \dot{\mathbf{x}}^{M}
\end{equation}
where $\mathbf{x}$ is the atmospheric state-vector, typically with $O(10^{10})$ degrees of freedom for NWP. Note that $\mathbf{x}^M$ represents the model state, whereas $\mathbf{x}^T$ will represent the true system state, in terms of model variables. 

Given the state $\mathbf{x}$ of a physical system, $M$ consists of all the known physics, forcings, and parameterizations of sub grid-scale processes. To make a one-timestep model forecast, we approximate the true change in state variables over that time by the model tendency, $\dot{\mathbf{x}}^{T} \approx M(\mathbf{x})$, and the error $\delta M$ in that approximation is called the \textsl{tendency error}:
\begin{equation}
\delta M \;=\; \dot{\mathbf{x}}^{T} - M(\mathbf{x})
\end{equation}

Even with perfect estimates of the current state of the atmosphere, the model tendency error would quickly separate forecasts from the truth, due to the atmosphere's chaotic dynamics \cite{lorenz1965}. As a result, techniques for reducing the model tendency error represent a current and active research area, and those that are applicable independent of the specific model are of special interest. Clearly, one would like to improve the physics represented by $M$ from first principles. In what follows, we assume this improvement has met the limit of diminishing returns, and move towards a statistical approach.

The general strategy of empirical correction is to compare short forecasts generated by a model to observations of the system being modeled over some training period. If the state-space of the system is well represented by the training period, and the model is a reasonable approximation of the true system, the forecast error statistics can be used to create an empirical correction that pushes the model closer to the truth \textsl{at each timestep} of numerical integration. Adjusting the model every timestep reduces the nonlinear growth of tendency error, providing more effective error reduction than a posteriori statistical correction \cite{danforth2008nonlinegrowth}. This strategy is similar to nudging or Newtonian relaxation in a data assimilation (DA) context, where one is assimilating observations, except that here we are nudging with predicted, rather than observed, forecast error.

The present study is an investigation of a three-step empirical correction procedure inspired by the work of Leith \cite{leith1978}, DelSole and Hou \cite{delsole1999}, and more recently Danforth et.\ al.\ \cite{danforth2007,danforth2008singularvals}. We first test its effectiveness in synchronizing Lorenz systems \cite{lorenz1963} with varied parameter-values in a perfect model scenario. We then apply the correction to an alternative model derived by Ehrhard and M\"uller \cite{ehrhard1990} tuned to approximate the evolution of a toroidal thermosyphon, an experimental analogue to the original Lorenz system. The ``true'' climate is represented by a long-time, high-dimensional computational fluid dynamics (CFD) simulation of the thermosyphon. An \textsl{analysis}, which is an approximation of the true system state in terms of model variables, is then created by three-dimensional variational (3DVar) data assimilation and used for training and verification of the empirical correction. This process mimics the application of empirical correction in an operational NWP setting. We also verify the corrected model by direct comparison with observations of the truth.

The results in each experiment suggest that the correction procedure is effective for reducing error. However, there is an associated cost of this short-term error reduction, which is evidenced by substantial qualitative differences between the dynamics of the corrected model and the true system, differences that were not present in the uncorrected model. Introduction of system-specific knowledge into the correction procedure is shown to mitigate some of that cost, while also improving error statistics further than the entirely general procedure.

The paper is structured as follows. In Sec.\ \ref{emcor} we define the procedure for a general model $M$.  The application of the technique in the perfect model scenario is addressed in Sec.\ \ref{pms}, and in Sec.\ \ref{tcm} we present the thermosyphon model correction.  Finally, we discuss the results and conclude the paper in Sec.\ \ref{conc}.

\section{\label{emcor}Empirical Correction}
The correction procedure employed in this experiment consists of three steps: (A) training, (B) state-independent correction and (C) state-dependent correction. The state-independent correction can be thought of as aligning the time-average of the model state with that of the true state. Likewise, the state-dependent correction can be considered an alignment of the model variance with that of the truth. To determine the correction terms, we compare short model forecasts to observations of the true system over a training period in a process called \textsl{direct insertion}. 

\subsection{\label{sec:di}Training}

In general, comparing model forecasts to a true physical system requires estimates of the true system state in terms of the model state-variables.  Consider a vector time-series $\mathbf{x}^{T}(t)$ of such estimates, which we will call the ``reference truth''. The amount of time $h$, measured in model timesteps, between estimates is called the \textsl{analysis window}; we assume it to be constant.

\begin{figure}[htp]%
\includegraphics[width=\columnwidth]{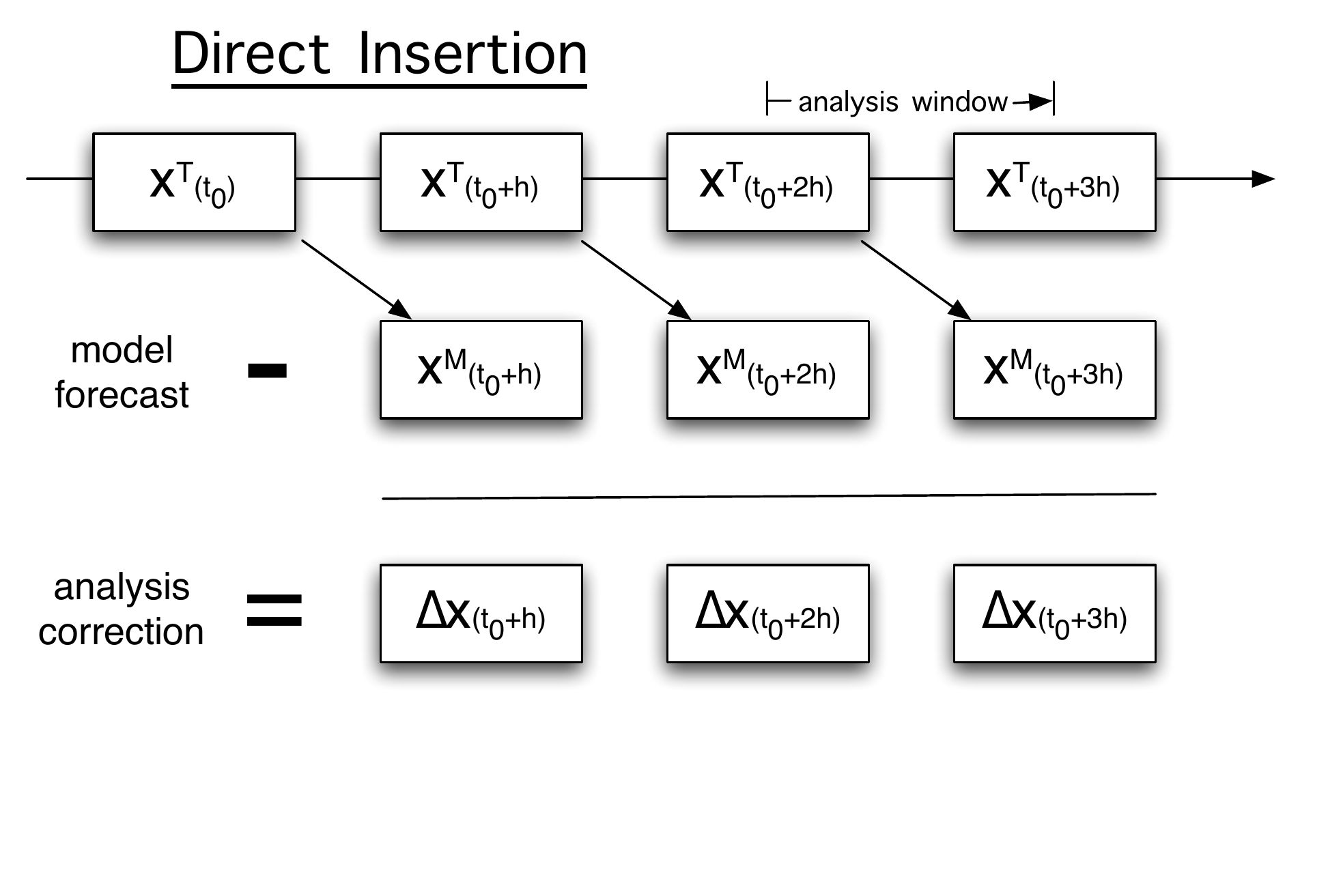} \vspace{-.6in}%
\caption{ The direct insertion procedure for comparing short model forecasts to the truth to obtain a time series of analysis corrections.
				$\mathbf{x}^T$ represents a time series of the reference truth, and the
				analysis window represents the number of timesteps between estimations of the true system state. 
				$\mathbf{x}^M$ represents a time series of forecasts with duration equal to the analysis window, each of which is started from the previous true state.
				The time-average of the analysis corrections $\langle \Delta \mathbf{x} \rangle$ divided by the number
				of timesteps in the analysis window $h$ approximates the average (state-independent) model bias $\mathbf{b}$.}%
\label{fig:dirins}%
\end{figure}

The process of direct insertion begins with the generation of a time-series $\mathbf{x}^{M}(t)$ of duration-$h$ model forecasts, where each forecast in the time-series is initialized from the previous state in the reference truth. The first vector in the series, for example, will be $\mathbf{x}^{M}(t_{0}+h)$, which is the model state resulting from an $h$-timestep forecast started with initial condition $\mathbf{x}^{T}(t_{0})$. Subtracting each of the model forecast states $\mathbf{x}^{M}(t)$ from the corresponding reference true state $\mathbf{x}^{T}(t)$ produces a third time-series $\Delta \mathbf{x}(t)$ which represents the forecast errors after $h$ timesteps. These errors result from differences between the model rate of change for each variable and the true rate of change, and they are commonly referred to as analysis corrections (or increments).  See Fig.\ \ref{fig:dirins} for a schematic of the procedure.

Finally, we separate each of the time series into anomalous and time-average components:
	\begin{align}
	\mathbf{x}^{T'}(t) \;&=\; \mathbf{x}^{T}(t) - \langle  \mathbf{x}^{T} \rangle \nonumber \\
	\mathbf{x}^{M'}(t) \;&=\; \mathbf{x}^{M}(t) - \langle  \mathbf{x}^{M} \rangle \label{eq:tsdecomp}\\
	\Delta \mathbf{x}'(t) \;&=\; \Delta \mathbf{x}(t) - \langle  \Delta \mathbf{x} \rangle \nonumber
	\end{align}
where the expectation operator $\langle\cdot\rangle$ denotes averaging over the training period, and the primes denote \textsl{anomalies}, which are differences from the mean.  The time-average components will be used for state-independent correction as described in Sec.\ \ref{sec:si} and the anomaly time-series will be used for state-dependent correction as detailed in Sec.\ \ref{sec:sd}.

\subsection{\label{sec:si}State-independent correction}

We turn our attention first to a state-independent correction of the form
\begin{equation}
	\dot{\mathbf{x}}^{T} \;\approx\; M^*(\mathbf{x}) \;\equiv\; M(\mathbf{x}) + \mathbf{b} 
\label{eq:sicortend1}
\end{equation}
where the constant vector $\mathbf{b}$ is the average model error (bias) to be determined. 

Recall that our goal here is to empirically align the time-averages of the state-variables in the model to those of the true system. We call the time-averaged true system state the \textsl{climatology}, and we approximate it by $\langle \mathbf{x}^T \rangle$, the average of the reference true state over the entire training period. The average of the analysis corrections $\langle\Delta \mathbf{x}\rangle$ over the training period provides an estimate for the systematic, state-independent error generated by the model during the analysis window, as explained in  Fig.\ \ref{fig:dirins}. Dividing by the number of timesteps in the analysis window, then, we approximate the model bias by $\mathbf{b} = \langle\Delta \mathbf{x}\rangle/h$, and the bias-corrected model tendency is thus given by
\begin{equation}
	M^*(\mathbf{x}) \;\equiv\; M(\mathbf{x}) + \frac{\langle \Delta \mathbf{x}\rangle}{h} 
\label{eq:sicortend2}
\end{equation}

Note that at this point we are approximating the model tendency error $\delta M$ by the model bias $\mathbf{b}$ alone. We also wish to estimate any component of error that may depend on the system state, by approximating $\delta M \approx \mathbf{b} + \mathbf{Lx}'$, where $\mathbf{L}$ is a matrix operator to be described in the next section. 

\subsection{\label{sec:sd}State-dependent correction}

To generate a linear state-dependent correction operator $\mathbf{L}$, we follow Leith \cite{leith1978} and DelSole and Hou \cite{delsole1999}, by first recomputing the forecast and correction time series in  Fig.\ \ref{fig:dirins} using the state-independent corrected model, $M^*$, and then decomposing into mean and anomalous components.  We seek an improved model of the form
	\begin{equation}
		\dot{\mathbf{x}}^{T} \;\approx\; M^+(\mathbf{x}) \;\equiv\; M^*(\mathbf{x}) + \mathbf{Lx}'
	\end{equation}
where $M^+$ includes both stages of correction. Letting $\mathbf{g} = \delta M^+ = \dot{\mathbf{x}}^{T} - [M^*(\mathbf{x}) + \mathbf{Lx}']$ be the tendency error of the improved model, we minimize the expected square tendency error $\langle \mathbf{g}^{tr}\mathbf{g}\rangle$, (where $\mathbf{g}^{tr}$ is the transpose of $\mathbf{g}$), with respect to \textbf{L}. The minimization results in the formula
	\begin{equation}
		\mathbf{L} \;=\; \langle \Delta \mathbf{x}'\mathbf{x}^{T'tr} \rangle \langle \mathbf{x}^{T'}\mathbf{x}^{T'tr} \rangle^{-1} 
		\;\equiv\; \mathbf{C}_{\Delta \mathbf{x}\mathbf{x}^T}\mathbf{C}_{\mathbf{x}^T\mathbf{x}^T}^{-1}
	\label{eq:Lcomp}
	\end{equation}
as explained by Danforth et.\ al.\ \cite{danforth2007}. $\mathbf{C}_{\Delta \mathbf{x}\mathbf{x}^T}$ is the average over the training sample of cross covariance matrices obtained by taking the outer product [$\Delta \mathbf{x}'(t)]\cdot[\mathbf{x}^{T'}(t)]^{tr}$ for each time $t$. $\mathbf{C}_{\mathbf{x}^T\mathbf{x}^T}$ is the average true-state covariance matrix, and \textbf{L} is known as Leith's state-dependent correction operator.

When \textbf{L} operates on the current anomalous state $\mathbf{x}'$, we can think of it as doing two things: (1) $\mathbf{C}_{\mathbf{x}^T\mathbf{x}^T}^{-1}$ operates on $\mathbf{x}'$, effectively relating the current state to the reference truth in the dependent sample, i.e. giving the best representation of the current state in terms of past states; and then (2) $\mathbf{C}_{\Delta \mathbf{x}\mathbf{x}^T}$ operates on the result, determining what correction should be made. This allows the model correction to adjust to different regions of state-space, and explains why the state-dependent correction can be thought of as attempting to align the model and true-state variances.  

In the next section, we describe the application of this three-stage procedure to align Lorenz systems with different parameter values in a perfect model scenario, and in Sec.\ \ref{tcm} we describe its application to couple a low dimensional model to a high dimensional toy climate simulation. We also note here that the term \textsl{corrected} will imply the application of both state-independent and state-dependent correction, unless explicitly stated otherwise.

\section{\label{pms}Perfect Model Scenario}
As a first step in the investigation of the correction technique, we consider its application to a model originally studied by Lorenz \cite{lorenz1963}. The system of equations (\ref{eq:lorenz}) represents fluid flow between two plates, Rayleigh-B\'enard convection, in which convection cells form for certain parameter ranges. However, with only slight modification (the details of which appear in Sec.\ \ref{tcmED}), they also describe the flow in a natural convection loop \cite{ehrhard1990,harris2011}. Lorenz systems are covered exhaustively in publication \cite{evans2004,yang2006}, and thus provide a familiar platform on which to perform preliminary tests of strategies for predicting the future state of chaotic systems.

\subsection{Experimental design}
\begin{figure}[htp]%
	\includegraphics[width=\columnwidth]{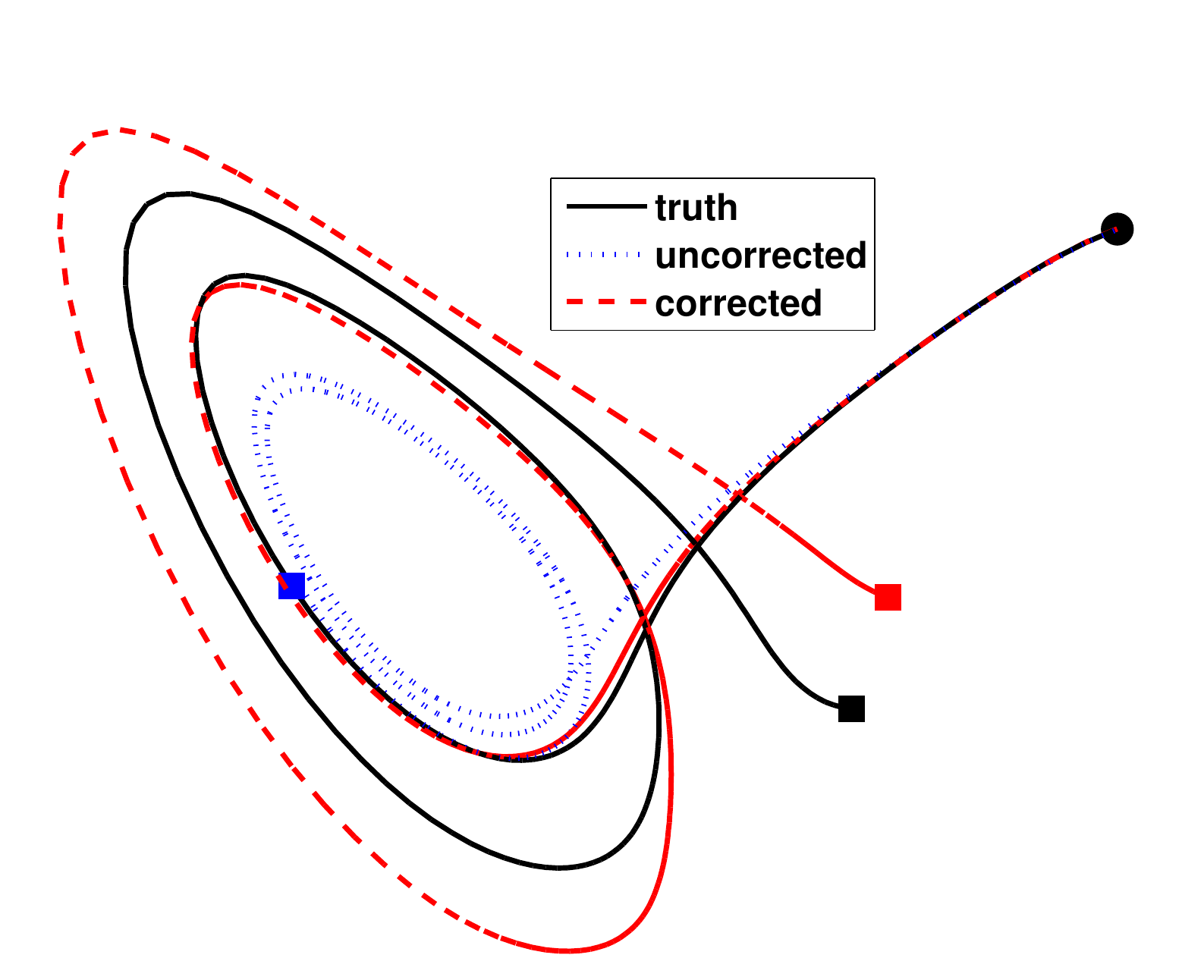} 
\caption{(Color online) Trajectories of the truth ($r=28$), a model with $r$-perturbation -2 ($r=26$), and the corrected model
				using an analysis window of $h=8$ timesteps. All start from the same initial condition (circle), and represent
				2 time-units (200 timesteps) of integration (concluding with squares). Note that the corrected model trajectory is well aligned with
				the true trajectory for much longer than the uncorrected model trajectory. Even after it deviates noticeably, the corrected model trajectory
				changes flow regimes (switches lobes) with the true trajectory. In contrast, the uncorrected model trajectory deviates
				from the true trajectory almost immediately, and remains in the initial lobe. }%
\label{fig:traj1}%
\end{figure}

In this perfect model scenario, the true system and the models share the structure of (\ref{eq:lorenz}) and only differ 
in parameter values. Specifically, a true or \textsl{nature} run was created by integrating the Lorenz system 
\begin{equation}
	\begin{array}{l}
	\dot{x}_1 \;=\; \sigma(x_2 - x_1) \\
	\dot{x}_2 \;=\; rx_1 - x_2 - x_1x_3 \\
	\dot{x}_3 \;=\; x_1x_2 - bx_3
	\end{array} 
	\label{eq:lorenz}
\end{equation}
with the standard parameter set: $\sigma=10,\; b=8/3,\; r=28$. Models with the same $\sigma$ and $b$, but with $r$-values varying from 25 to 31 in increments of 0.5 (except for $r=28$) were the subjects for correction. For each of these 12 models, the correction algorithm was performed using the 4 different analysis windows $h = 1, 2, 4,$ and 8 timesteps, resulting in 48 distinct model-correction pairs in an exponential design. The training and testing of the corrected models is detailed in the following sections, and a picture showing one particularly positive outcome appears in Fig.\ \ref{fig:traj1}.

\subsection{Training}
A 100 time-unit nature run was generated by integrating system (\ref{eq:lorenz}) from the initial condition $\mathbf{x}_0=[1.508870, \; -1.531271, \;\, 25.46091]^{tr}$ (see \cite{miller1994}) for 10000 timesteps of $\kappa = 0.01$ time units each, using fourth order Runge-Kutta. For each of the 12 models and 4 analysis windows, a time-series of short forecasts was generated by direct insertion (see Fig.\ \ref{fig:dirins}). As an illustrative example, consider training with an analysis window of $h=4$.  The first forecast, $\mathbf{x}^{M}(t_{0}+4)$, is a 4-timestep model forecast started with the true initial condition $\mathbf{x}^{T}(t_{0}) = \mathbf{x}_{0}$. The second forecast, $\mathbf{x}^{M}(t_{0}+8)$, is a 4-timestep model forecast started with the true state $\mathbf{x}^{T}(t_{0} + 4)$, and so on, resulting in 10000/4 = 2500 total short forecasts. The state-independent correction $\langle\Delta \mathbf{x}\rangle/h$ was then computed as described in Fig.\ \ref{fig:dirins} and Sec.\ \ref{sec:si}.  

Next, the correction time-series $\Delta \mathbf{x}(t)$ was recomputed using a state-independent corrected version of the model, namely Eq.\ (\ref{eq:sicortend2}). Correction and true-state anomalies were calculated as in Eq.\ (\ref{eq:tsdecomp}), and the cross covariance and covariance matrices were determined for each time $t$ and averaged over the training sample to obtain $\mathbf{C}_{\Delta \mathbf{x}\mathbf{x}^T}$ and $\mathbf{C}_{\mathbf{x}^T\mathbf{x}^T}$ as outlined in Sec.\ \ref{sec:sd}. Finally, the static Leith operator \textbf{L} was computed and the training procedure was complete. 

Note that the training design imparts a statistical disadvantage upon the use of wider analysis windows. Specifically, doubling the analysis window halves the number of samples in the training period. The design was chosen, despite this prejudice, to more accurately reflect an operational implementation in which the training data is likely to be drawn from a fixed period of time. However, to further support the validity of comparisons between models corrected with different analysis windows, we note that letting the training period be $10000h$, ensuring that the number of samples is held constant at 10000, yields results that are qualitatively indistinguishable from those presented here.


\begin{figure}[htc]
     \centering 
     \includegraphics[width=.95\columnwidth]{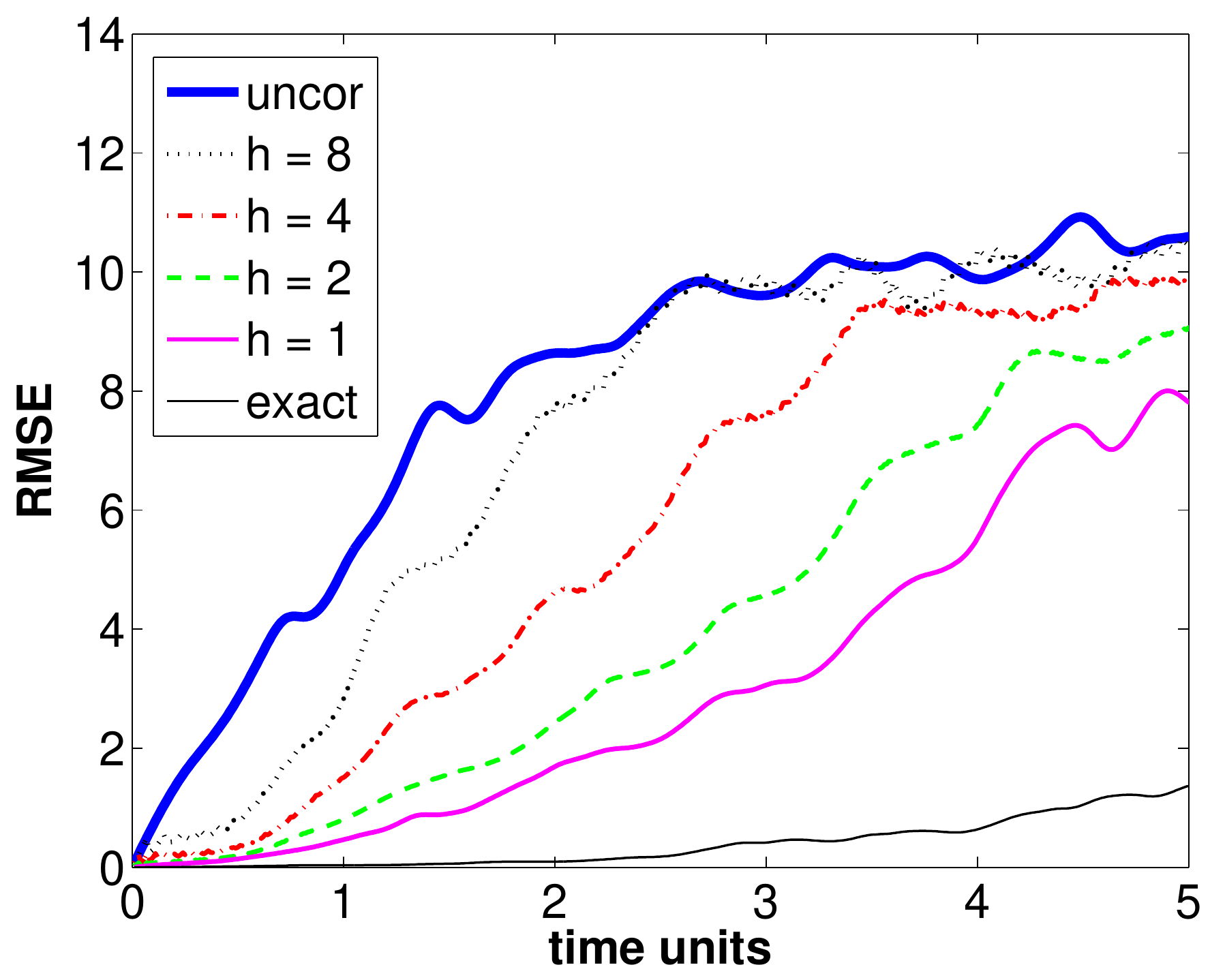}
     \caption{\label{fig:allrmse} (Color online) Plots of average RMSE over 1000 trials for the uncorrected (thick solid blue) and corrected models (all with $r=26$) using analysis 				windows of 8, 4, 2 and 1, black dot,  red dash-dot, green dash, solid magenta respectively.  The exact model (thin solid black) is the same as the true system, but started from an initial condition
     	perturbed randomly on the order of $10^{-3}$ in each state variable. Time units are on the $x$-axis, where
     	one time unit is 100 timesteps in the numerical integration. The technique's performance clearly degrades with widening analysis window.}  
	\end{figure}


\subsection{Testing}
A new nature run, 10000 time-units (one million timesteps) in length, was generated starting from the last true state in the training period. The purpose of beginning at the end of the training period was to obtain an \textsl{independent} truth with which to test the effectiveness of the corrected models. For each of the 48 corrected models, 1000 randomly selected states from this new nature run were used as the initial state, and both the uncorrected and corrected models were integrated for 20 time units. In addition, an \textsl{exact} model, with the same parameters as the truth, was integrated from a random perturbation of that same initial condition, on the order of $10^{-3}$ in each state variable. The purpose of the exact model is to represent an upper bound for the effectiveness of the empirical model error correction, i.e. demonstrating a case where the impact of the $r$-perturbation was corrected exactly, and forecast error comes only from the initial condition discrepancy. 

Fig.\ \ref{fig:traj1} depicts the results of a single test with a particularly positive outcome. Trajectories of the truth (solid black), uncorrected model (blue dot), and corrected model (red dash), all start from the same initial state and progress for two time-units. The corrected model trajectory stays close to the true trajectory for much longer than the uncorrected model trajectory, and even after deviating markedly, the corrected model trajectory still switches lobes with the true trajectory.

Two metrics were used to measure forecast accuracy: root mean square error (RMSE), and anomaly correlation (AC). The RMSE is given at time $t$ by
\begin{equation}
	\text{RMSE}(t) \;=\; \sqrt{\frac{1}{N}\displaystyle\sum_{n=1}^N{\left[x_n^T(t)-x_n^M(t)\right]^2}}
\label{eq:test.1}
\end{equation}
where $\mathbf{x}^M$ is the model state and $\mathbf{x}^T$ is the true state, and we are summing over the $N=3$ state variables in the Lorenz system. Fig.\ \ref{fig:allrmse} plots the average RMSE over 1000 trials performed for correction of the $r=26$ ($r$-perturbation of -2) model using the 4 different analysis windows.

\begin{figure}[htp]
     \centering
     \subfigure{
          \label{fig:acbyaw}
          \includegraphics[width=\columnwidth]{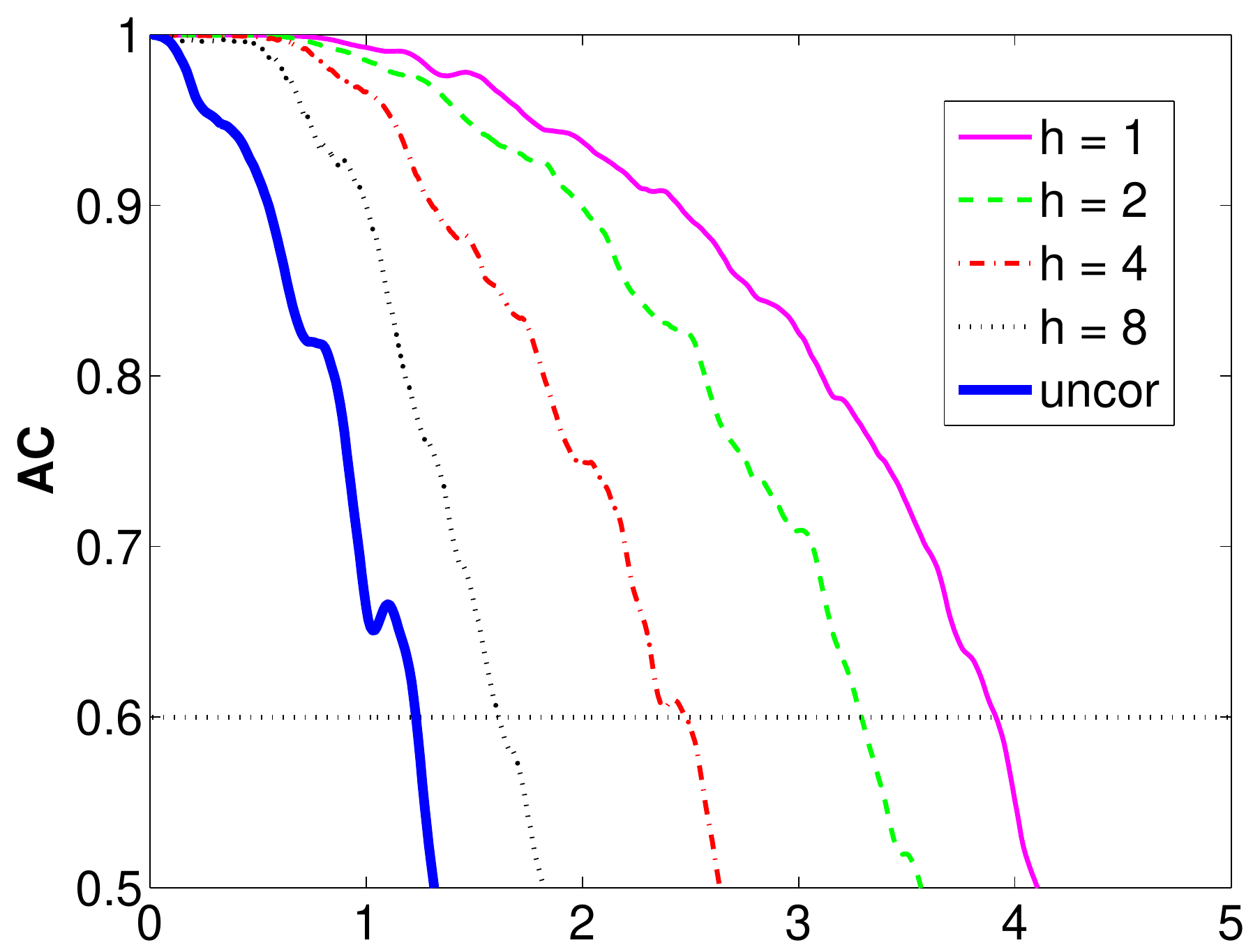}} \\
     \subfigure{
          \label{fig:acbyrpert}
          \includegraphics[width=\columnwidth]{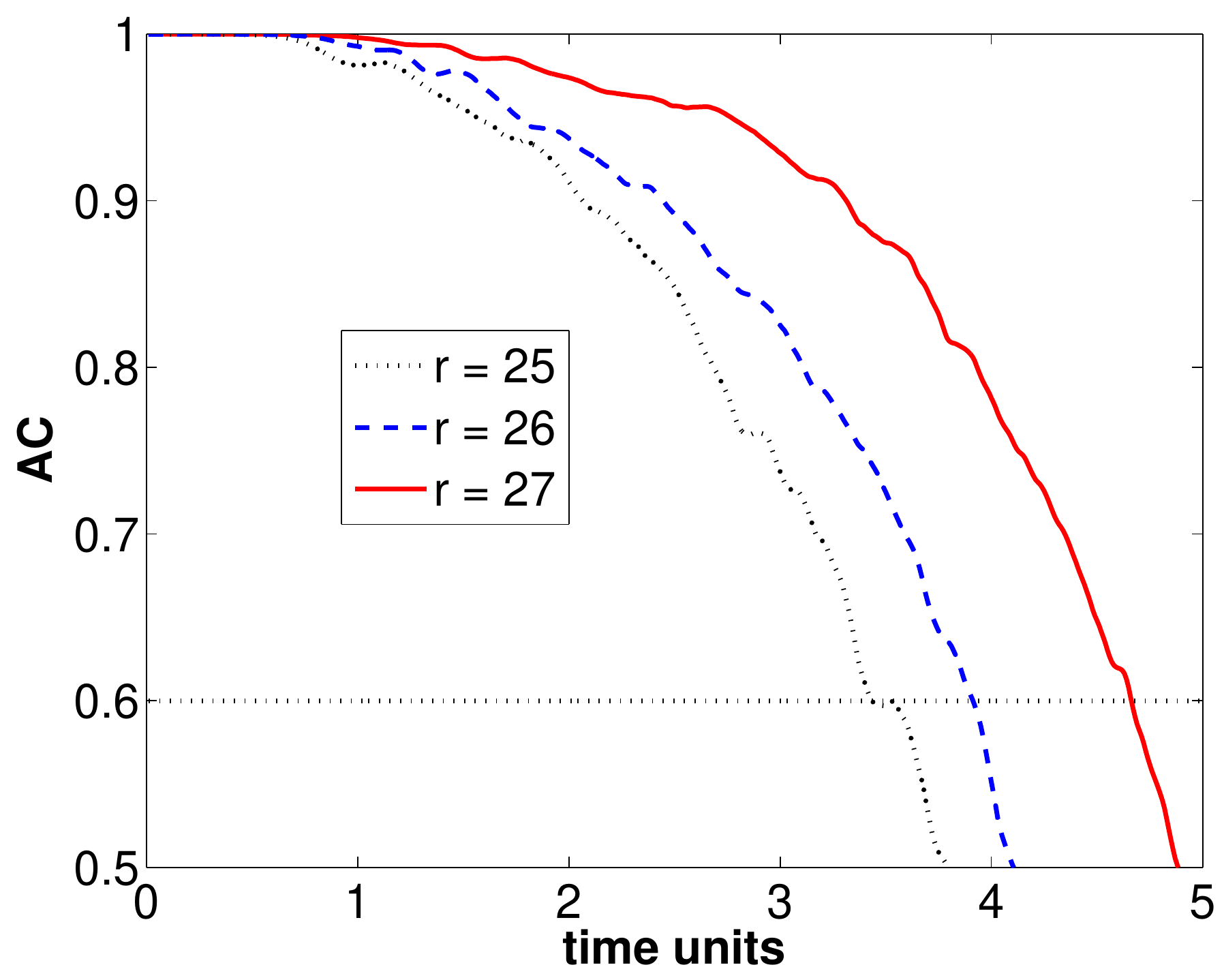}}
     \caption{\label{fig:acbyraw} (Color online) (top) Plots of average AC over 1000 trials for the uncorrected (thick solid blue) and corrected models (all with $r=26$) using analysis windows of $h = 8$, 4, 2 and 1, black dot,  red dash-dot, green dash, solid magenta respectively. In agreement with Fig.\ \ref{fig:allrmse}, performance drops as the analysis 		
     	window widens. Note, however, that even with $h = 4$ timesteps the corrected model provides a useful forecast for twice as long as the 
     	uncorrected model. (bottom) Plots of average AC for models corrected with an analysis window of 1 timestep, and $r = $ 25, 26 and 27, black dot, blue dash, and 
     	solid red respectively. The greater the magnitude of $r$-perturbation, the more difficult it is to correct successfully. Similar results are observed for $r$-values greater than 28.}  
	\end{figure}

Anomaly correlation is a metric frequently used in weather and climate modeling to determine the length of time for which a model forecast is useful. The AC is given by 
\begin{equation}
	\text{AC} \;=\; \frac{ \mathbf{x}^{M'} \cdot \mathbf{x}^{T'} }{ ||\mathbf{x}^{M'}||_2||\mathbf{x}^{T'}||_2 }
\label{eq:test.3}
\end{equation}
where $\mathbf{x}^{M'}$ and $\mathbf{x}^{T'}$ are the anomalous model state and anomalous true state, respectively, at a particular time. AC is essentially the dot product of the anomalous model state with the anomalous true state, normalized such that AC $= 1$ for a perfect model. A forecast is typically considered useful for as long as its AC remains above 0.6 \cite{kalnaybook2003}.
As with RMSE, the AC scores for each model, corrected and uncorrected, were averaged over 1000 trials to provide a good representation of model performance. See Fig.\ \ref{fig:acbyraw} for AC plots demonstrating the effects of changing analysis window length and parameter perturbation in the original model on the duration of useful forecasts.

\subsection{Results}
Fig.\ \ref{fig:allrmse} demonstrates that the tested empirical correction technique succeeds in reducing error in the perfect model scenario. However, the importance of frequent observations of the truth during the training period is highlighted by the approach of the average corrected model RMSE towards the uncorrected model RMSE with widening of the analysis window. It is also noteworthy that the corrected model remains almost as good as the exact model for a full time unit when observations are made every timestep. In this case the RMSE remains below 2 more than 5 times as long the uncorrected model. 

Fig.\ \ref{fig:acbyraw} demonstrates that along with reduced error, empirical correction has the potential to provide forecasts that are useful for much longer. Training with an analysis window of 1 timestep, the corrected model forecasts are useful for nearly 4 times longer than the uncorrected model forecasts. In light of the sensitivity of AC to analysis window length, the bottom panel of Fig.\ \ref{fig:acbyraw} suggests that the accuracy of parameter values matters less for the effectiveness of the corrected model, as measured by error statistics, than does the frequency of observations in training. However, it should be noted that error statistics are not the whole story. The ability of the corrected models to reproduce the qualitative dynamical \textsl{behavior} of the true system accurately, like switching of lobes in the attractor, is only indirectly indicated by reduced forecast error. We address this issue in Sec.\ \ref{sec:mqd}. 

\begin{figure}[htp]
     \centering
     \subfigure{
          \label{fig:skdurbyaw}
          \includegraphics[width=.9\columnwidth]{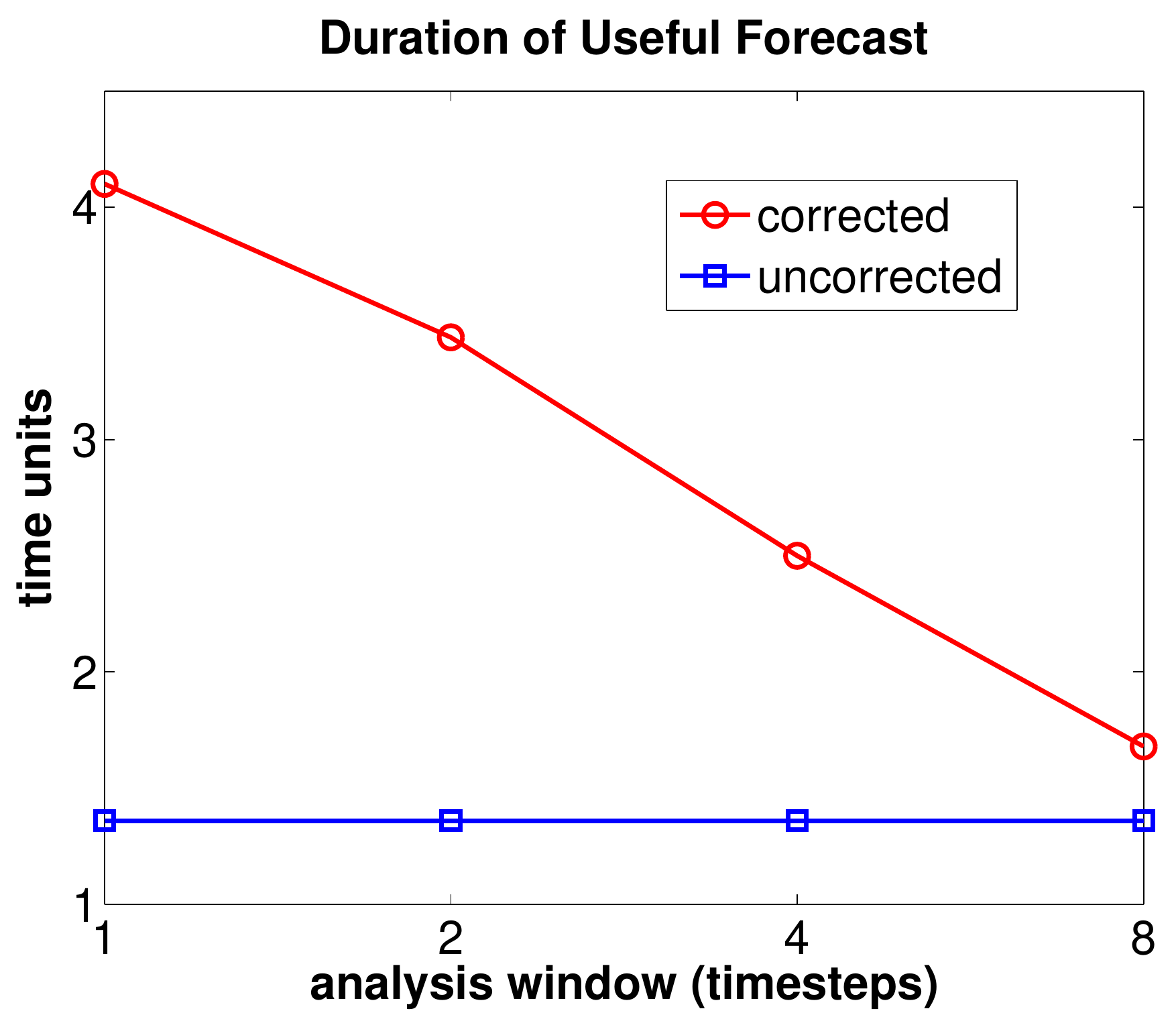}} \\
     \subfigure{
          \label{fig:skdurbyrpert} 
          \includegraphics[width=.9\columnwidth]{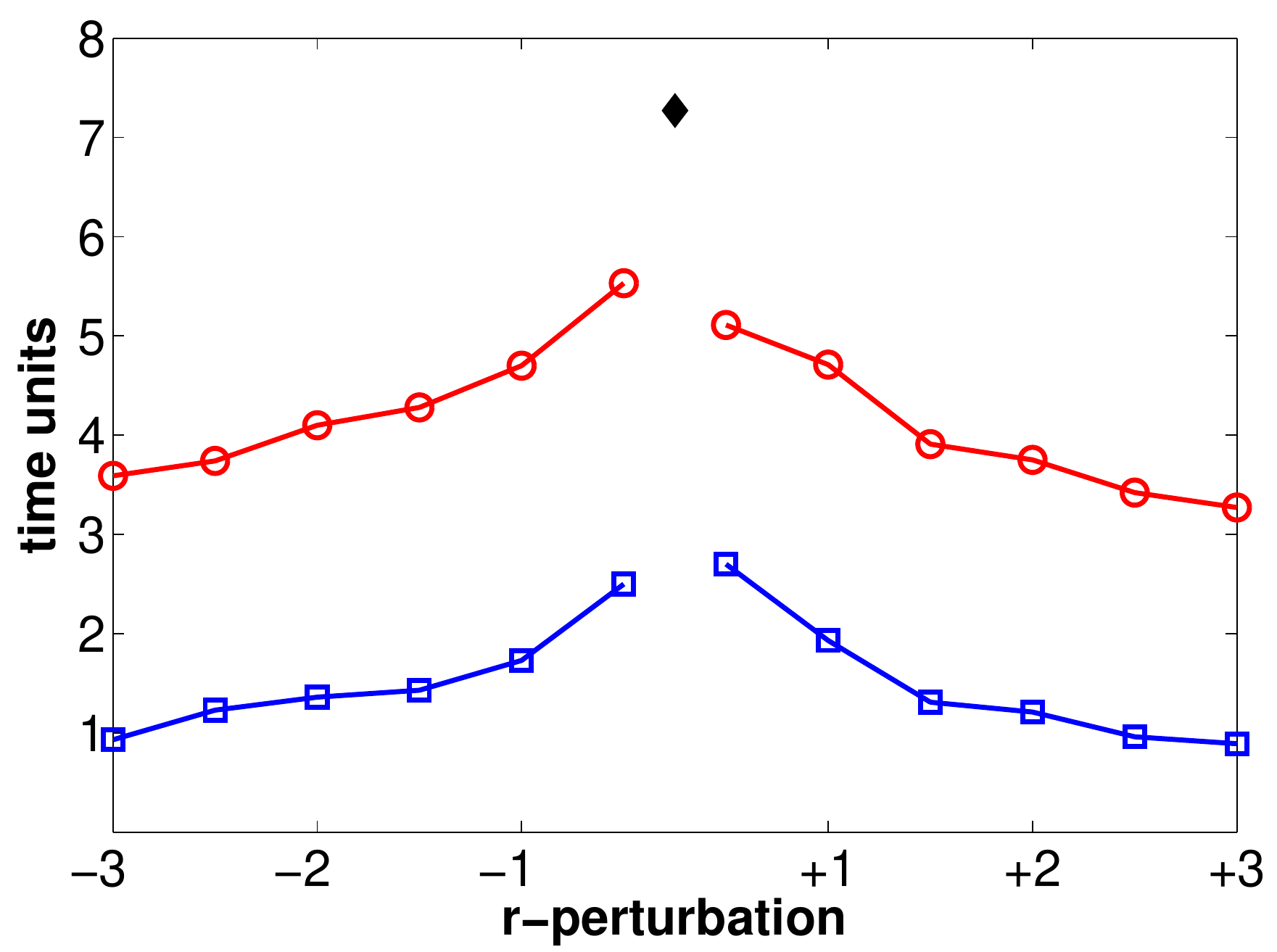}}
     \caption{\label{fig:skdurbyraw} (Color online) Plots of the duration of useful forecasts (first time AC=0.6) vs.\ analysis window (top) and
     					$r$-perturbation (bottom). (top) The drop in duration of useful skill with analysis window suggests
     					an asymptotic decrease in the value of correction towards 
     					no improvement from the uncorrected $r=26$ model. (bottom) Using an analysis window of 1 timestep, we plot the duration as a function of $r$-perturbation. The black diamond is the average duration of a useful forecast for the exact model (true system with perturbed initial state), representing the limit of predictability imposed by initial condition uncertainty. Note: the corrected models with greatest $r$-perturbation outperform the uncorrected models with least $r$-perturbation.}  
	\end{figure}	

\begin{figure}[htp]%
\includegraphics[width=\columnwidth]{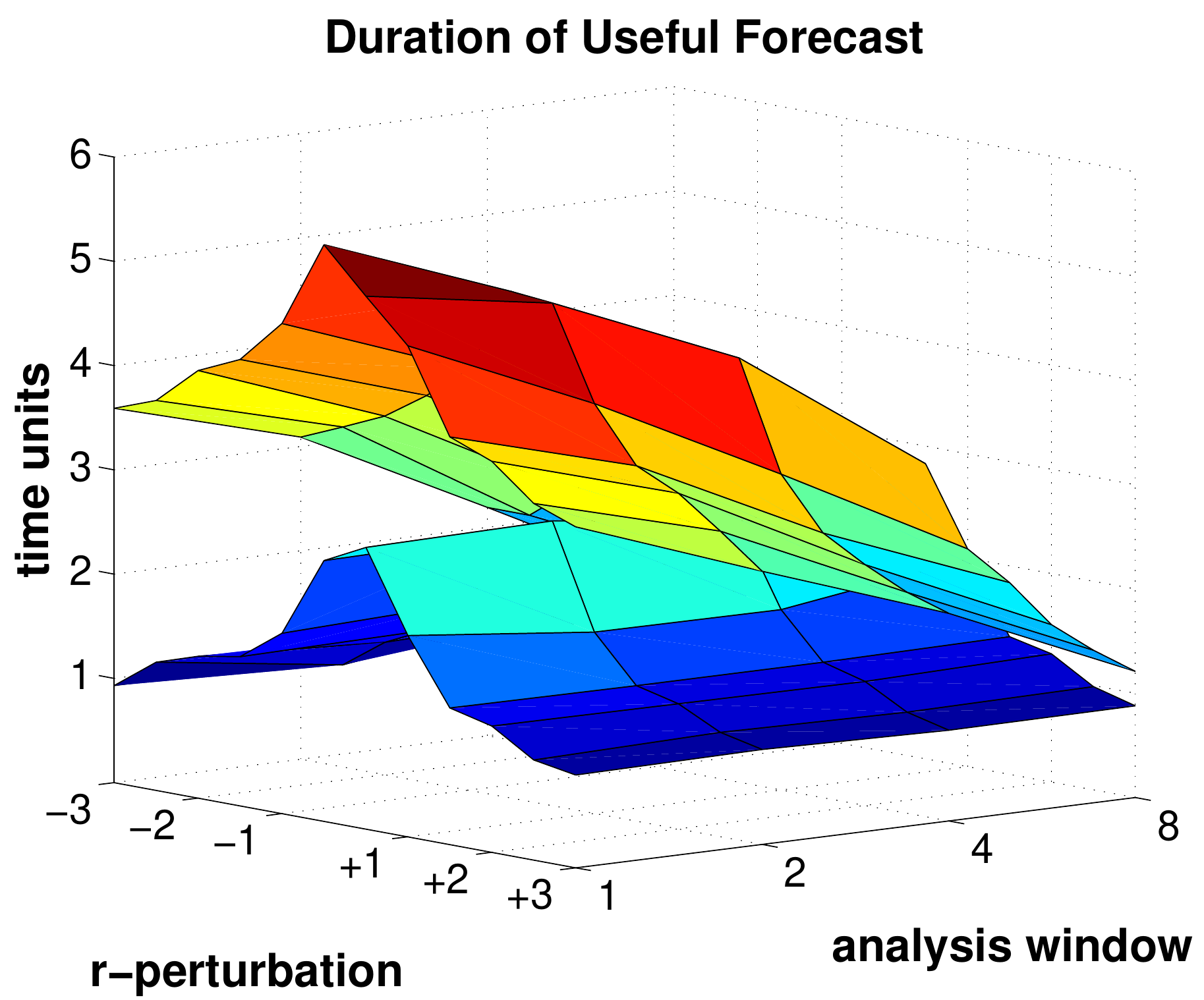}\vspace{-.1in}\\%
\caption{(Color online) Surface plot showing duration of useful forecasts for all combinations of analysis window and $r$-perturbation tested, top surface corrected, bottom surface uncorrected. On the left face
we can see the cross section represented by Fig.\ \ref{fig:skdurbyrpert}. Using an analysis window of 2 timesteps, corrected models with largest r-perturbation provide forecasts that are useful longer than those of uncorrected models with smallest r-perturbation.}%
\label{fig:skillful}%
\end{figure}

A summary representation of the data in Fig.\ \ref{fig:acbyraw} is shown in Fig.\ \ref{fig:skdurbyraw}, where the duration of forecast usefulness is plotted vs.\ analysis window in the top panel, and vs.\ $r$-perturbation in the bottom panel.  However, these plots still focus only on two cross sections of the 48 total model-correction pairs that were tested in the perfect model scenario, specifically the 4 corrected $r=26$ models in the top and the 12 corrections across all $r$-perturbations using an analysis window of 1 timestep in the bottom. To get a better sense of the relative importance, Fig.\ \ref{fig:skillful} shows a surface plot of duration of useful forecasts for all combinations of analysis window and $r$-perturbation tested. Surprisingly, with an analysis window of 2 timesteps, the corrected models with the largest r-perturbation (parameter error greater than 10\%) produce forecasts that are useful longer than those made by the uncorrected model with the smallest r-perturbation (parameter error less than 2\%). For systems with reasonably small model errors, this is an indication that empirical correction may improve forecasts more readily than parameter tuning.

\section{\label{tcm}Toy Climate Model}
\begin{figure}[htp]%
\includegraphics[width=\columnwidth]{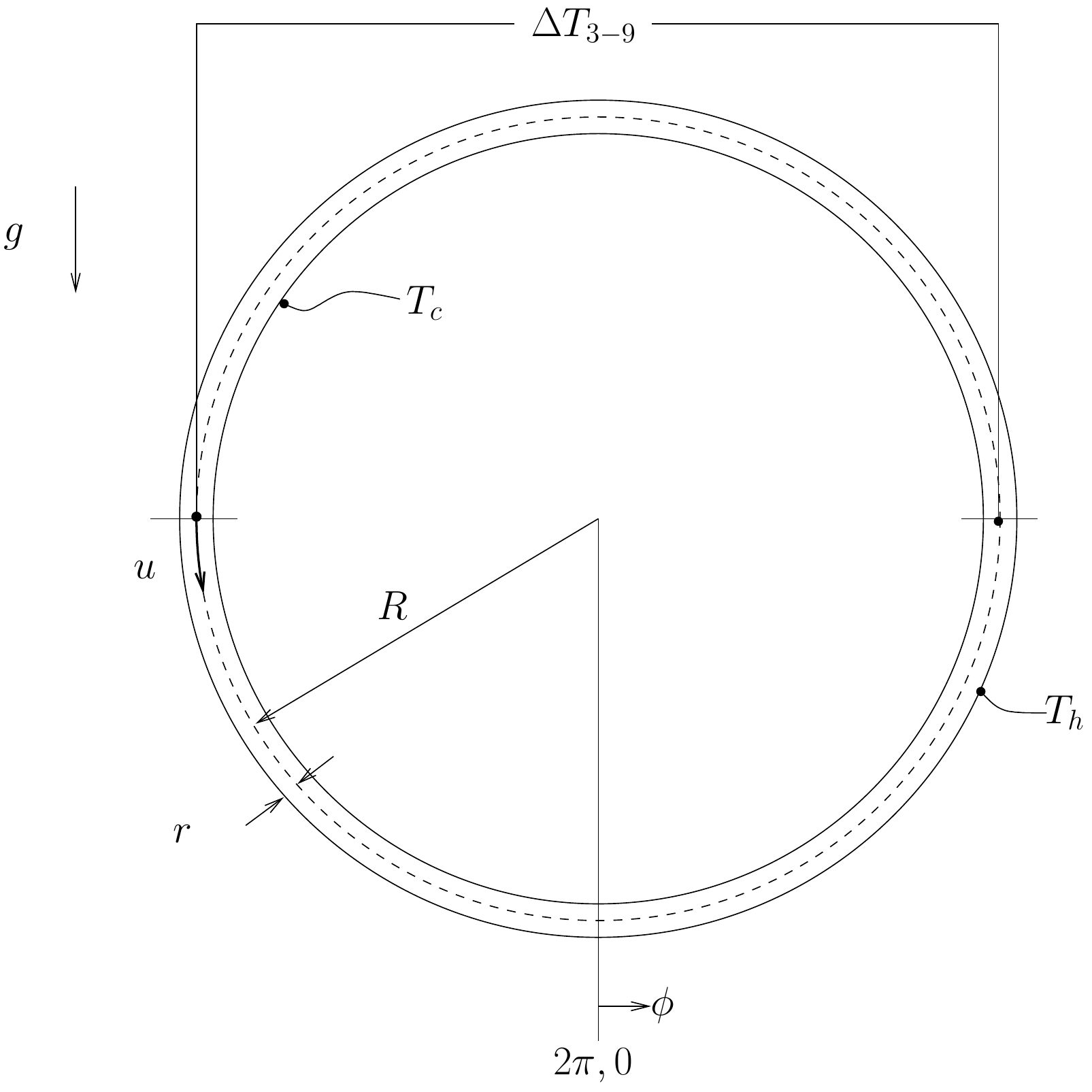} \vspace{-.3in} \\%
\caption{ Schematic of the natural convection loop, or thermosyphon.  The top half of the fluid-filled loop is cooled to the temperature $T_{c}$, while the bottom half is heated to $T_{h} > T_{c}$, creating a temperature inversion.  The three state variables in the low-dimensional model are (1) $x_{1} \propto \bar{u}$, the mass flow rate (mean fluid velocity); (2) $x_{2} \propto \Delta T_{3-9}$, the horizontal temperature difference; and (3) $x_{3} \propto \Delta T_{6-12} - \Delta T^{\text{cond.}}_{6-12}$, the difference between the vertical temperature profile and the value it takes during conduction.}%
\label{fig:loop}%
\end{figure}

We next investigate the effectiveness of the correction procedure in a more realistic situation, where the forecast model is structurally different (in dynamics, dimension, parameterization, etc...) from the true system.  Consider a fluid-filled, vertically-oriented natural convection loop, or thermosyphon, with circular geometry, a schematic of which appears in Fig.\ \ref{fig:loop}. The constant temperature imposed on the wall of the lower half of the loop, $T_{h}$, is greater than the constant temperature imposed on the wall of the upper half, $T_{c}$, resulting in a temperature inversion. For large enough temperature differences $\Delta T_{w} = T_{h}-T_{c}$, convection dominates, and the flow undergoes chaotic reversals of direction referred to as \textsl{flow regime changes}, while remaining laminar \cite{ehrhard1990,harris2011}.  These dynamics produce forecasting difficulties similar to those encountered in weather and climate prediction, and thus the thermosyphon provides a useful platform on which to test potential improvements to forecasting methods. 

\subsection{\label{tcmED}Experimental design}
The true system was represented by numerical simulation using the 2-D laminar Navier-Stokes equations along with the energy equation, 
and a finite-volume-based flow modeling software package (FLUENT 6.3) was used to perform the numerical integration, see \cite{harris2011} for details. Almost 90 days of fluid behavior was generated, in which $O(10^4)$ flow reversals occurred. We also note that for the Rayleigh number used in this experiment, Ra $ = 1.5\times10^5$, the thermosyphon has two unstable convective equilibrium solutions corresponding to steady clockwise and steady counter-clockwise flow \cite{ridouane2010,harris2011}.

The low-dimensional model used to make forecasts of the CFD simulation is the Ehrhard-M\"uller (EM) system:
\begin{equation}
	\begin{array}{l}
	\dot{x}_1 \;=\; \alpha(x_2 - x_1) \\
	\dot{x}_2 \;=\; \beta x_1 - x_2(1 + KH(|x_{1}|)) - x_1x_3 \\
	\dot{x}_3 \;=\; x_1x_2 - x_3(1 + KH(|x_{1}|))
	\end{array} 
	\label{eq:em}	
\end{equation}
where $x_{1}$ is proportional to the mean fluid velocity, $x_{2}$ is proportional to the horizontal temperature difference in the loop, and $x_{3}$ is proportional to the deviation of the vertical temperature difference from the value it takes during conduction. This system was derived from physical principles to model a natural convection loop \cite{ehrhard1990,harris2011}, and for this study the parameters $\alpha = 7, \beta = 33, K = 0.07$ were tuned empirically to best match the flow reversal behavior of the CFD simulated thermosyphon. 

The primary difference between EM and Lorenz systems is $H$, a function that determines the velocity dependence of the heat transfer between the fluid and the wall. This characteristic of the flow is ignored by the Lorenz equations (i.e.\ $K=0$). $H$ varies as the third root of the magnitude of the mean fluid velocity for $|x_{1}|>1$, and as a fourth degree polynomial in $|x_{1}|$ for $|x_1|\le1$ to remain differentiable; the reader may see \cite{harris2011} for more detail. We note that when $K=0$ in the EM equations (\ref{eq:em}), they are identical to the Lorenz system (\ref{eq:lorenz}) with $b=1$. Physically, the unitary geometric factor (i.e.\ $b=1$) in EM results from the forced single circular convection cell in a thermosyphon, as opposed to the unconstrained flow producing multiple cells between two plates. 

Using a background forecast created with the EM model, and observations of the CFD mean fluid velocity $\bar{u}$ with Gaussian noise added to simulate error, 3DVar data assimilation was performed to generate an analysis, or best guess of the true state of the system in terms of the variables of the forecast model \cite{harris2011}.  One segment of the analysis is used as the reference truth $\mathbf{x}^{T}(t)$ for training, and the remainder is used for testing.   

\subsection{Training}
Approximately 3 days of 3DVar analysis, corresponding to 432 time-units in the EM forecast model, was used as the training period reference truth. As in the perfect model scenario, fourth-order Runge-Kutta was used with a timestep of $\kappa = 0.01$ time-units. An analysis window of $h =5$ timesteps, corresponding to about 30 seconds of simulated flow, was used to match the frequency of observation in the data assimilation scheme. Thus, 8640 short forecasts were used to compute the model bias and Leith operator for the model, as outlined in Sec.\ \ref{emcor}.

\subsection{Testing}
Three forecast models were compared by verification against both the 3DVar analysis and direct observation of the mass flow-rate $\bar{u}$ in the CFD simulated thermosyphon: (1) the uncorrected EM model with parameters tuned to best represent observations of the CFD simulated mass flow rate; (2) the tuned model with correction applied; and (3) an EM model whose parameters differ from the tuned model by 10\%, \textsl{with} correction applied. The purpose of the third test-model is to gauge the relative capabilities of empirical correction and parameter tuning for error reduction and prolonging the usefulness of forecasts. 

\begin{figure}[htp]
     \centering
     \subfigure{
          \label{fig:emcorAC}
          \includegraphics[width=.9\columnwidth]{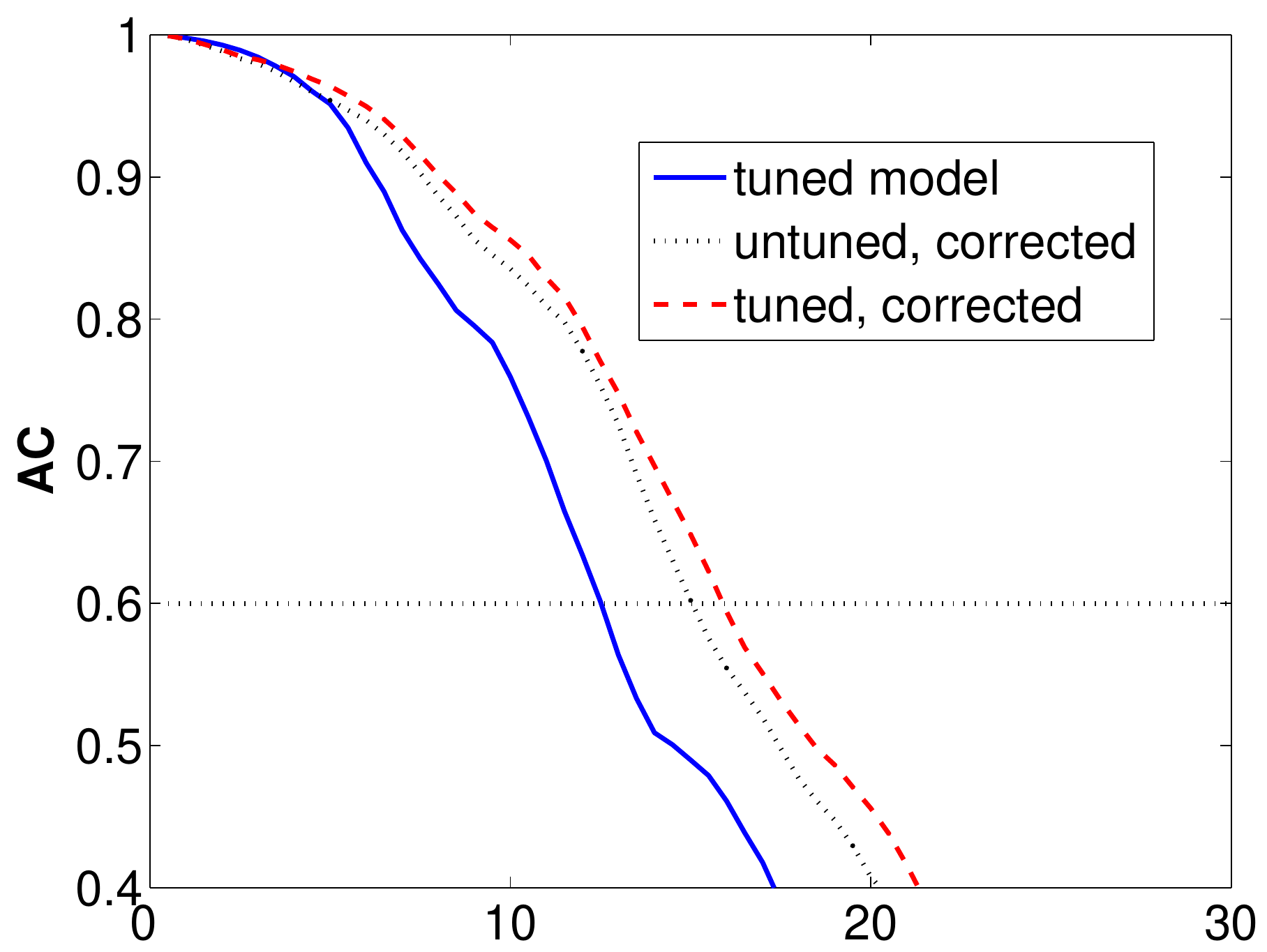}} \\
     \subfigure{
          \label{fig:emcorMFRE}
          \includegraphics[width=.9\columnwidth]{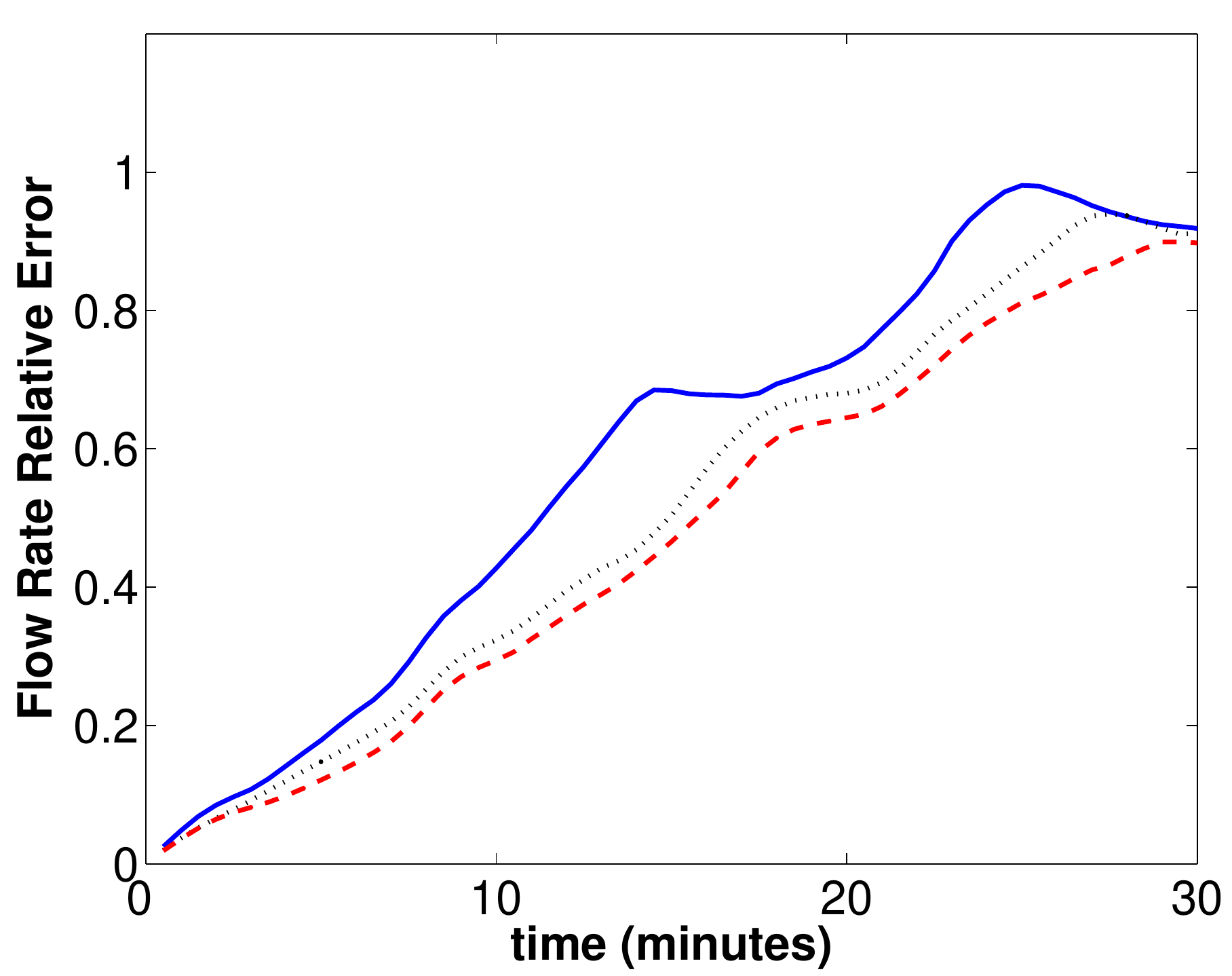}}
     \caption{\label{fig:emcor} (Color online) Plots showing verification of the corrected model with respect to (top) 3DVar analysis and (bottom) direct observation of the CFD simulated mass flow-rate $\bar{u}$. (top) We see that empirical correction produces forecasts that are useful approximately 25\% longer than forecasts of the uncorrected model. Also, correction of an untuned model (10\% difference from optimal value in every parameter) produces forecasts that are useful far longer than the uncorrected but optimally tuned model. (bottom) The corrected model shows reduced error when verified against direct observations of the mean fluid velocity in the simulation as well. Again we see that correction of a model with sub-optimal parameterization produces better results than tuning the parameters optimally but neglecting to apply empirical correction.  }  
\end{figure}

A set of 1000 trial forecasts were performed, starting from randomly chosen points in the analysis \textsl{after} the training period. For each forecast, RMSE and AC time-series were computed with respect to the analysis and averaged over all trials.  See the top panel of Fig.\ \ref{fig:emcor} for the resulting average AC plot (RMSE not shown). We also verify the model forecasts against observed scalar mass flow-rate $\bar{u}$, for which time-series of relative error were averaged over the 1000 trials, pictured in the bottom panel of Fig.\ \ref{fig:emcor}. 

Two details are important in the computation of the relative error.  First, to compare model output to observations, it is necessary to convert the model state-variables to ``observation-space'' variables. In other words, an observation operator determined by data assimilation was used to convert the model state-vector $(x_{1},x_{2},x_{3})^{tr}$ to an observation-space value $\tilde{x}_{1}$, which is the predicted mass flow-rate of the system. Second, the error is taken relative to the saturation point, which we define as the average absolute difference between the mass flow-rates of the system at randomly chosen points in time. Thus, an average relative error near 1 means that the forecast model is no better than a random guess.

The results presented in Fig.\ \ref{fig:emcor} indicate that corrected models, tuned or not, produce smaller short-term forecast error on average than the uncorrected, optimally tuned model. Corrected model forecasts are thus typically useful for longer. Though this is an important benefit, average short-term error statistics may conceal considerable qualitative differences between model dynamics and those underlying the true system. Stability of equilibrium solutions, and changes of flow regime characterized by aperiodic switching between otherwise confined regions of state-space are examples of qualitative characteristics for which it may be crucial that the model dynamics match those of the truth. In the next section we address the effect of empirical correction on the dynamical matching capability of the EM forecast model.

\subsection{\label{sec:mqd}Matching qualitative dynamics}

\begin{figure}[htp]
     \centering
     \subfigure{
          \label{fig:rcdiffbyicuncor}
          \includegraphics[width=.9\columnwidth]{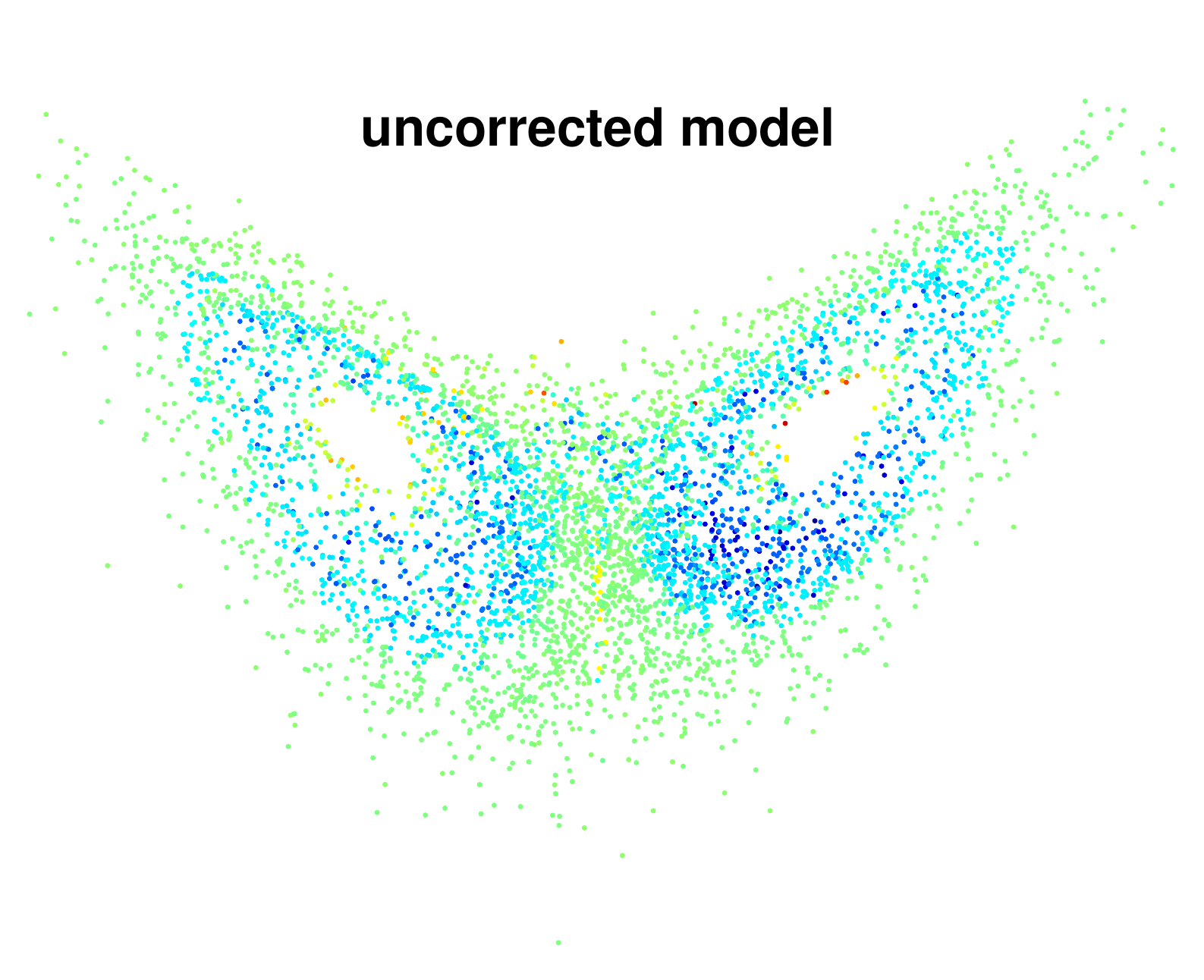}}  \\
     \vspace{-.4in} 
     \subfigure{
          \label{fig:rcdiffbyiccor}
          \includegraphics[width=.9\columnwidth]{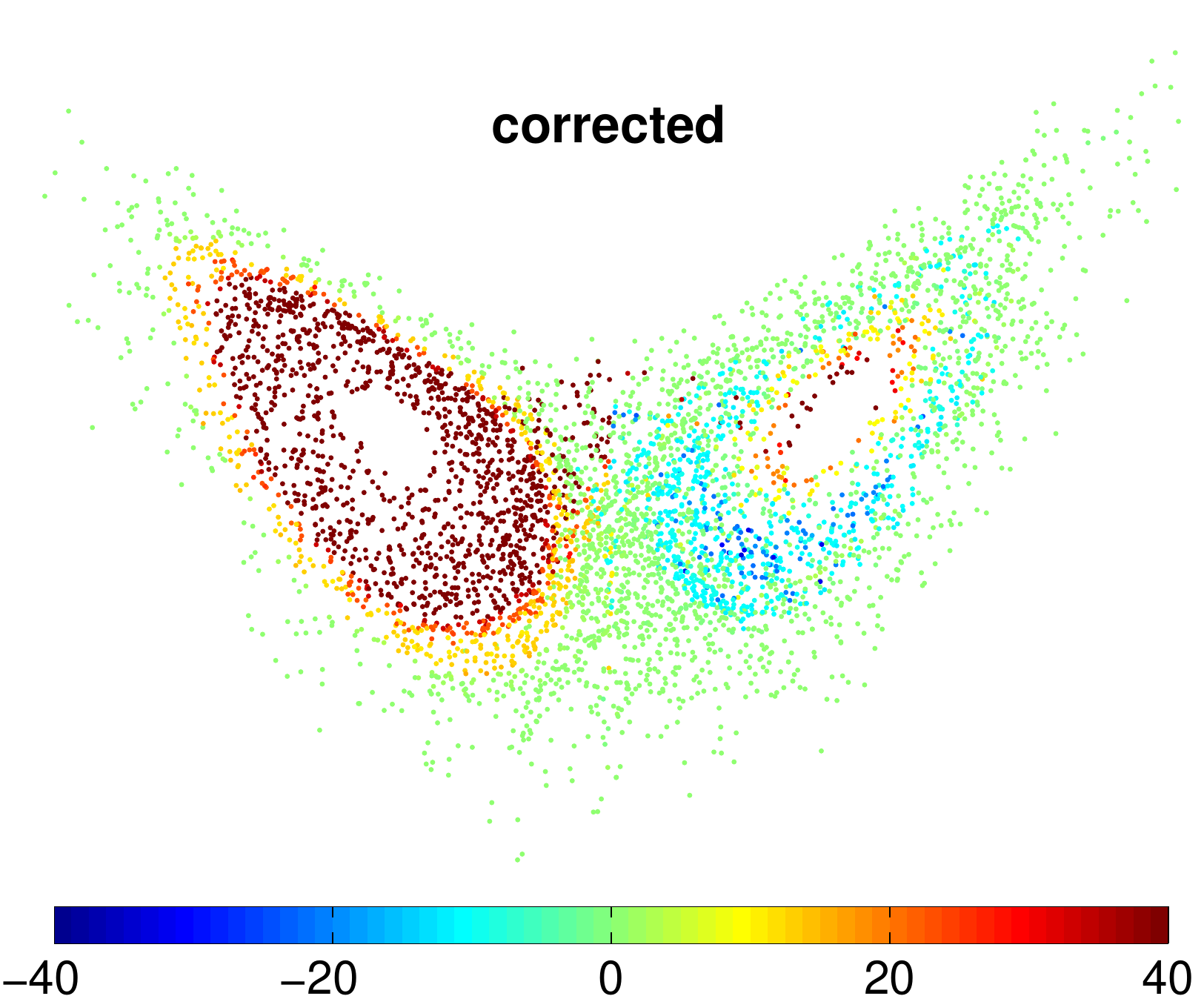}} \\
     \caption{\label{fig:rcdiffbyic}  (Color online) Difference (in minutes) between predicted and actual time of first flow reversal, plotted by initial state, for the uncorrected (top) and corrected (bottom) models. The difference was taken $t_{\text{model}} - t_{\text{actual}}$, so that positive values (towards red) indicate late predictions while negative values (towards blue) indicate early predictions. First, note that for initial states near the edge of the attractor, in either lobe, each model performs well in predicting the coming flow reversal. This makes sense because for these initial states the change of flow regimes is imminent, and thus easily predicted. Next, observe that when the uncorrected model errs, it almost always makes an early prediction (blue dots). Model trajectories close to the unstable convecting equilibrium solution at the center of each lobe oscillate with larger amplitude than in the true system. One might expect that empirical correction would counteract the effect, and in fact that is apparent (bottom). However, rather than improving flow reversal prediction, the results here indicate extreme overcorrection. The dark red dots are not 40-minute late predictions; the limit of the color axis was chosen to better illustrate the spread in the remaining data. In actuality, the corrected model predicts that a flow reversal \textsl{will never happen} for those initial states. Finally, observe that the empirical correction has manufactured a lobe asymmetry, characterized by the much smaller region of dark red in the right lobe. }  
\end{figure}

To measure the accuracy of models with regard to matching the flow reversal behavior of the true system, forecasts were generated with both the corrected and uncorrected EM models for 5000 initial states throughout the attractor, from the testing portion of the 90-day 3DVar analysis, and the time of the first predicted flow reversal was recorded for each one. We investigate the difference between the predicted times and the actual times (from the analysis) of first flow reversal, taken $t_{\text{model}} - t_{\text{actual}}$, so that positive values indicate late predictions while negative values indicate early predictions. See Fig.\ \ref{fig:rcdiffbyic} for plots of the results.  

Three unforeseen costs of empirical correction for this system, in terms of lost dynamical matching capability, can be summarized as follows:
\begin{itemize}
	\item[1)] Stabilization of convective equilibrium solutions
	\item[2)] Elimination of flow reversal behavior for states in a neighborhood of either convective equilibrium
	\item[3)] Spurious dynamical asymmetry between lobes
\end{itemize}
We address these costs in the sections that follow, examining the first two related phenomena in Sec.\ \ref{sec:eqstab}, and then explaining the spurious asymmetry in Sec.\ \ref{sec:brokensym}.

\subsubsection{\label{sec:eqstab} Stabilization of equilibrium solutions}

Jacobian analysis of the EM equations (\ref{eq:em}) provides analytical confirmation of the instability of the convective equilibria in the uncorrected model. Specifically, the Jacobian evaluated at each equilibrium has one negative real eigenvalue, whose eigenvectors are in the local direction of (tangent to) the stable manifold of the equilibrium, and a conjugate pair of complex eigenvalues with positive real part, whose 2-D eigenspace is locally tangent to the unstable manifold of the equilibrium. In fact, for both convective equilibria the positive real parts of these unstable eigenvalues are quite small, on the order of $10^{-2}$, indicating weakly repelling instability. In the following we explain analytically the mechanism by which empirical correction overcomes this weak repulsion, producing a forecast model with \textsl{attracting}, and thus stable, convective equilibria, see Fig.\ \ref{fig:falsestab}.

Empirical correction of the EM model effectively alters the right-hand side of the differential equations (\ref{eq:em}) by first adding a constant related to the bias term $\mathbf{b}$, and then adding a term that depends linearly on the model state, i.e., something related to $\mathbf{Lx}'$. Letting $\mathbf{f}^M$ be the vector-valued EM differential equation, and $\mathbf{f}^{M^+}$ be the corrected equation, we write
\begin{equation}
  \mathbf{f}^{M^+} \;=\; \mathbf{f}^M + \mathbf{b}_0 + \mathbf{L}_0\mathbf{x}'
\end{equation} 
where
\begin{equation}
  \mathbf{b}_0 = \lim_{\kappa\to0}\mathbf{b} \;\;\;\;\;\;\;\;\text{and}\;\;\;\;\;\;\;\; \mathbf{L}_0 = \lim_{\kappa\to0}\mathbf{L}
\end{equation}
can be thought of as the computed bias term and Leith operator, respectively, for an infinitesimal timestep $\kappa$.  Because of the nonlinearity of the system, we cannot determine the exact relationship between the infinitesimal correctors $\mathbf{b}_0$ and $\mathbf{L}_0$, and the bias term $\mathbf{b}$ and Leith operator $\mathbf{L}$, respectively, that we compute using a timestep of $\kappa = 0.01$ and analysis window of $h=5$ timesteps. However, since we discretize $\mathbf{f}^{M^+}$ in numerical integration, we \textsl{can} determine the $\mathbf{b}_0$ and $\mathbf{L}_0$ that we actually apply. We effectively approximate the correction terms
\begin{equation}
  \mathbf{b}_0 \approx \frac{\mathbf{b}}{\kappa} = 100\mathbf{b} \;\;\;\;\;\;\;\;\text{and}\;\;\;\;\;\;\;\; \mathbf{L}_0 \approx \frac{\mathbf{L}}{\kappa} = 100\mathbf{L}
\end{equation}
within the fourth order Runge-Kutta scheme.

Now, armed with an analytical representation of the differential equations $\mathbf{f}^{M^+}$, we note the following relationship between the corrected model Jacobian $D\mathbf{f}^{M^+}$, and that of the uncorrected EM model $D\mathbf{f}^M$:
\begin{equation}
  D\mathbf{f}^{M^+} \;=\; D\mathbf{f}^M \;+\; \mathbf{L}_0 \;\approx\; D\mathbf{f}^M \;+\; 100\mathbf{L}
\end{equation}
since the constant bias term $\mathbf{b}_0$ disappears and $\mathbf{L}_0$ operates on a translation of the model state. We evaluate the Jacobian of the corrected model at each of the convective equilibria, and determine its eigenvalues. Indeed, the real part of the complex conjugate eigenvalues, for each convective equilibria, is \textsl{negative} for the corrected model $\mathbf{f}^{M^+}$, and thus the convective equilibria have become stable.

The equilibria attract all states inside neighborhoods around them, which thereby separate state-space near the attractor into regions whose trajectories will either change flow regime at least once, or not at all. See Fig.\ \ref{fig:falsestab} for an example of this effect. Furthermore, any trajectories that land in one of these neighborhoods after only one or two flow reversals, which might occur within the expected 17-minute duration of useful forecasts for the corrected model reported in Fig.\ \ref{fig:emcor}, will then approach steady convection. This behavior is in qualitative \textsl{opposition} to that of the true system, \textsl{and} that of the original uncorrected EM model, for which steady convection in a single direction is an unstable equilibrium.

\subsubsection{\label{sec:brokensym} Broken symmetry}

The size discrepancy between left and right-lobe regions attracted to the convective equlibria of the corrected model revealed in Fig.\ \ref{fig:rcdiffbyic} demonstrates that empirical correction breaks the symmetry of the EM system. As in the conventional Lorenz system (\ref{eq:lorenz}), the EM system (\ref{eq:em}) is symmetric under the mapping $(x_1,x_2,x_3)\mapsto(-x_1,-x_2,x_3)$. Again letting $\mathbf{f}^M$ be the vector-valued EM differential equation, this symmetry implies that $\mathbf{f}^M$ commutes with a certain matrix $A$, i.e.
\begin{equation}
	A\mathbf{f}^M(\mathbf{x}) \;=\; \mathbf{f}^M(A\mathbf{x}), \;\;\;\;\; A \;=\; \left[\begin{array}{rrr} -1&0&0\\0&-1&0\\0&0&\;\;\,1\end{array}\right]
\end{equation}
In fact, empirical correction breaks this symmetry in two ways. 

First, recall that after bias correction alone we have changed the EM system by adding a constant vector to the right-hand side of the differential equation. Letting $\mathbf{f}^{M^*}$ represent the bias-corrected differential equation, $\mathbf{f}^{M^*}$ \textsl{does not} commute with $A$ unless there is zero bias in $x_1$ and $x_2$, i.e.
\begin{equation}
	A\mathbf{f}^{M^*}(\mathbf{x}) \;\ne\; \mathbf{f}^{M^*}(A\mathbf{x}) \;\;\;\;\forall\,\mathbf{b}_0\ne\left[\begin{array}{l}0\\0\\b_3\end{array}\right]
\end{equation}
where $b_3$ can be any constant, and we recall that $\mathbf{b}_0 = \lim_{\kappa\to0}\mathbf{b}$ can be thought of as the computed bias term for an infinitesimal timestep $\kappa$. Note that even if no bias in $x_1$ or $x_2$ existed, the probability of statistically computing a bias term $\mathbf{b}$ that would preserve the symmetry of the EM system is zero.

In the unlikely case that a bias term is computed that preserves symmetry, or such a bias term is forced, state-dependent correction will break it. Assuming that $\mathbf{f}^{M^*}$ does commute with $A$, and letting $\mathbf{f}^{M^+}$ be the fully corrected differential equation, then
\begin{equation}
	A\mathbf{f}^{M^+}(\mathbf{x}) \;=\; \mathbf{f}^{M^+}(A\mathbf{x}) \;\;\Longleftrightarrow\;\; A\mathbf{L}_0 \;=\; \mathbf{L}_0 A
\end{equation}
In other words, $\mathbf{f}^{M^+}$ commutes with $A$ if and only if $\mathbf{L}_0=\lim_{\kappa\to0}\mathbf{L}$ commutes with $A$. This forces the computed $\mathbf{L}$ to be of the form
\begin{equation}
	\mathbf{L} \;=\; \left[\begin{array}{lll}\ell_{11}&\ell_{12}&0\\\ell_{21}&\ell_{22}&0\\0&0&\ell_{33}\end{array}\right]
\end{equation}
where the $\ell_{ij}$ can be any constants. Of course the probability of computing such an $\mathbf{L}$ statistically is also zero. Therefore, both bias and state-dependent correction break the symmetry of the EM model.

\subsubsection{Summary}

Undesirable effects of empirical correction include the breaking of system symmetry, and altered stability of equilibrium solutions which results in large regions of state-space for which flow-reversals never occur in corrected model forecasts. We emphasize that the qualitative behavior of the corrected model is substantively different from that of the uncorrected model, which matches the behavior of the CFD simulated truth. This is true despite the fact that the corrected model shows improved average error statistics. However, it is possible to adjust the correction procedure to mitigate this effect by directly incorporating dynamical knowledge of the true system, which is the subject of the next section. Note that in doing so, we sacrifice the general applicability of the technique.

\begin{figure}[htp]%
	\includegraphics[width=\columnwidth]{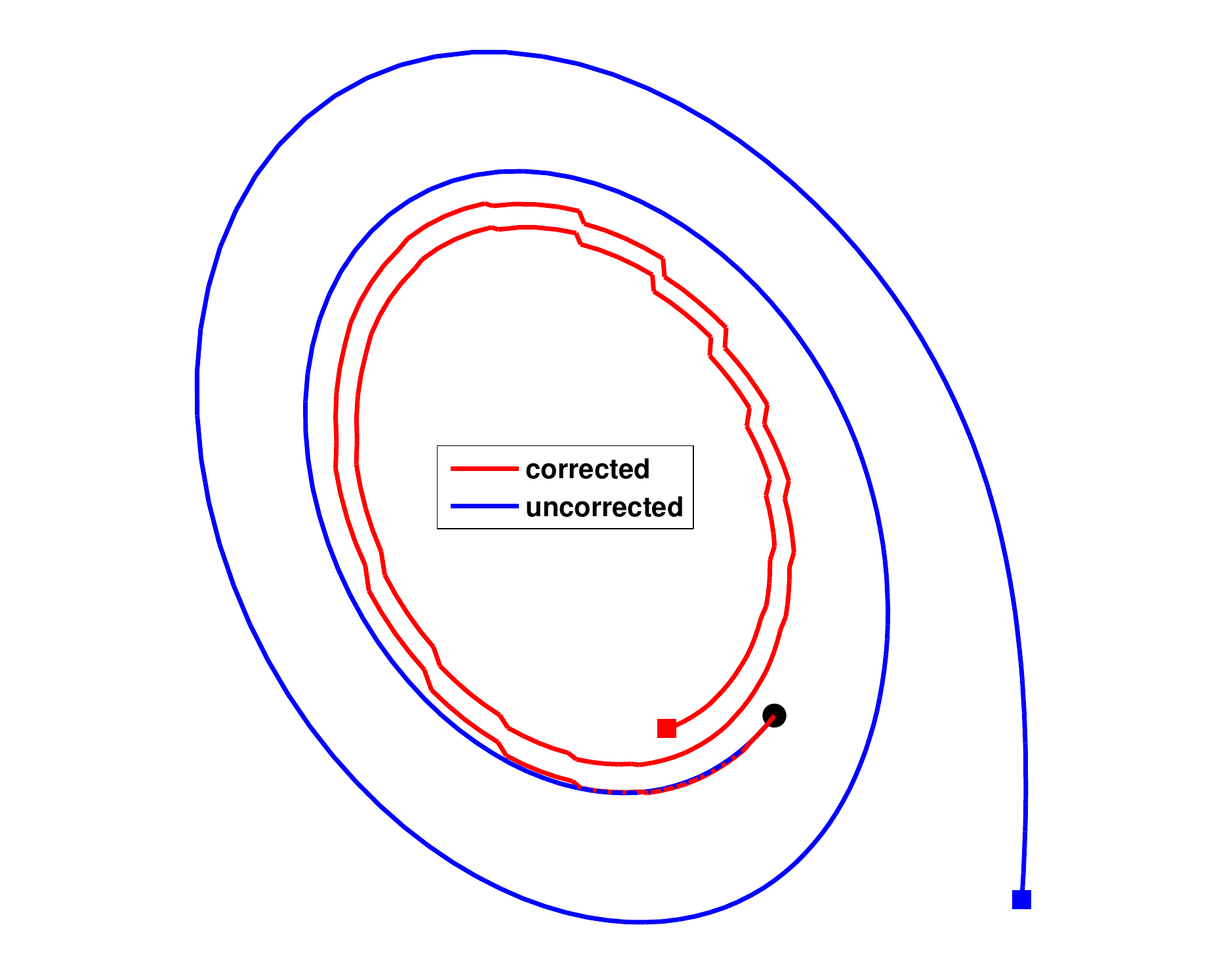} 
\caption{ (Color online) The first 20 minutes of a forecast for an initial state attracted to the left-lobe convective equilibrium stabalized by the empirical correction (i.e., one of the dark red dots towards the center of the left lobe in the bottom of Fig.\ \ref{fig:rcdiffbyic}). The uncorrected EM model trajectory (blue) behaves as it should, winding away from the equilibrium at the center of the lobe, whereas the trajectory of the corrected model (red) collapses toward the equilibrium solution, indicating the false stability produced by the correction. The source of the false stability is the overcorrective nudging visible as jumps (every 30 seconds, the length of the analysis window) on the red curve. The correction is attempting to lengthen flow regimes, a consequence of the blue regions near the equilibria in the top of Fig.\ \ref{fig:rcdiffbyic}. }%
\label{fig:falsestab}%
\end{figure}

\subsection{\label{sec:idk}Incorporating dynamical knowledge}

To encode dynamical knowledge of the system in the empirical correction procedure, the state-space is partitioned into regions based on the qualitative behavior of the system, and then a separate bias term $\mathbf{b}$ and state-dependent correction operator $\mathbf{L}$ is computed \textsl{for each region}. For example, in the context of weather forecasting, the state-space of the atmosphere could be divided by stage in the El Ni\~no oscillation, or by day and night, or local season for regional models. Fig.\ \ref{fig:rcdiffbyic} suggests two ways to partition the state-space in the present context: (1) by flow regime direction (lobe); or (2) by distance from the nearest convective equilibrium solution. In each case, the state-space is decomposed into two regions, left/right lobe, or near/far from equilibrium, respectively. In addition to testing the correction procedure using each of these strategies individually, a procedure applying them simultaneously, which results in a partition of the state-space into four regions, is also tested. 

\subsubsection{Lobe dependence}

To generate lobe-dependent bias correction terms and state-dependent correction operators, two regions $L_{1,2}$, corresponding to flow regimes of opposite direction, are defined by
\begin{equation}
	L_1 \;=\; \{\mathbf{x}: x_1 < 0\}, \;\;\;\;\;\;\;\; L_2 \;=\; \{\mathbf{x}: x_1 \ge 0\}
	\label{eq:lobes}
\end{equation}
noting that their union is the entire state-space. Physically, $L_1 \; (L_2)$ represents all states undergoing clockwise (counter-clockwise) convection. Next, the direct insertion procedure illustrated in Fig.\ \ref{fig:dirins} is modified to produce two subsequences of the analysis correction time series, $\Delta\mathbf{x}_{L_1}$ and $\Delta\mathbf{x}_{L_2}$, that correspond to two subsequences of the analysis time series, $\mathbf{x}^T_{L_1}$ and $\mathbf{x}^T_{L_2}$, respectively, where
\begin{equation}
  	\mathbf{x}^T_{L_1} = \left\{\mathbf{x}^T(t): \mathbf{x}^T(t) \in L_{1}\right\}
\end{equation}
and $\mathbf{x}^T_{L_{2}}$ is defined similarly. These subsequences are separated into mean and anomalous components, as in Eqs.\ (\ref{eq:tsdecomp}), using means over each individual subsequence. Finally, the separate correction terms and operators are computed by:
\begin{align}
  \mathbf{b}_{L_k} &= \frac{\langle\Delta \mathbf{x}_{L_k}\rangle}{h}, \;\;\;\;\;\;\;
  \mathbf{L}_{L_k} = \mathbf{C}_{\Delta\mathbf{x}_{L_k}\mathbf{x}^T_{L_{k}}}\mathbf{C}^{-1}_{\mathbf{x}^T_{L_k}\mathbf{x}^T_{L_{k}}} \label{eq:Lkcorterms} 
\end{align}
for $k=1,2$. To apply the lobe-dependent correction, at every timestep of numerical integration the current state is determined to be in either $L_1$ or $L_2$, and the appropriate bias correction term and state-dependent operator are applied to advance the model.
\\
\subsubsection{Dependence on distance from equilibrium}

A procedure analogous to that for lobe-dependent correction is applied here. Two regions $E_{1,2}$, corresponding to near and far from equilibrium, respectively, are defined by
\begin{equation}
  \begin{array}{l}
	E_1 \;=\; \{\mathbf{x}: \text{min}(||\mathbf{x} - \mathbf{eq_1}||_2, ||\mathbf{x} - \mathbf{eq_2}||_2) < c_v \},  \\ 
	E_2 \;=\; \{\mathbf{x}: \mathbf{x}\notin E_1\}
	\end{array}
	\label{eq:equis}
\end{equation}
where $\mathbf{eq_1}$ and $\mathbf{eq_2}$ are the convective equilibria (estimated from the parameters of uncorrected model, \cite{harris2011}), and the critical value $c_v$ is a parameter for the procedure. We are effectively approximating the neighborhoods attracted to the convective equilibria by spheres of radius $c_v$. For all results shown in the paper, the critical value $c_v=8.5$ was used, though error statistics and dynamical matching capability were virtually unaffected by changing this parameter by 25\% in either direction. This range of critical values was tested as estimates to the average radius of the dark red region in the left lobe of Fig.\ \ref{fig:rcdiffbyic} (bottom). Continuing with the correction scheme, direct insertion is modified as in the lobe-dependent correction, and the region-specific bias terms and state-dependent operators are calculated as in Eqs.\ (\ref{eq:Lkcorterms}), substituting $E_k$ for $L_k$. Application of the correction to a forecast model also proceeds in the same fashion.

\subsubsection{Simultaneous lobe and equilibrium dependence} 

Defining the lobe regions $L_{1,2}$ as in the lobe-dependent section, and the equilibrium regions $E_{1,2}$ as in the previous section, we define the four regions for simultaneous lobe and equilibrium-dependent correction by
\begin{equation}
  \begin{array}{l}
  R_1 \;=\; L_1 \cap E_1, \;\;\;\;\;\; R_2 \;=\; L_1 \cap E_2 \\
  R_3 \;=\; L_2 \cap E_1, \;\;\;\;\;\; R_4 \;=\; L_2 \cap E_2
  \end{array}
\end{equation}
so that we modify direct insertion to produce four subsequences of analysis increments, each paired with the appropriate subsequence of the analysis time series. Note that the critical value $c_v$ used in defining the regions $E_{1,2}$ does not depend on the lobe in this scheme. Allowing a different $c_v$ for each lobe is a possible modification that was not tested. We compute the bias terms and Leith operators as in Eqs.\ (\ref{eq:Lkcorterms}), substituting $R_k$ for $L_k$, where now $k=1,2,3,4$. Again, application of the correction proceeds as in the individual dynamically informed schemes, where the current state is determined to be in one of the four defined regions, and the appropriate bias term and Leith operator are used to advance the model.  

\subsection{Results}

First, we discuss the results of applying the dynamically uninformed, and entirely general correction procedure as described in Sec.\ \ref{emcor}, to couple the three-dimensional EM forecast model to the high-dimensional CFD simulated thermosyphon. Then we report the results of modifying the correction procedure to directly incorporate dynamical knowledge of the system, as detailed in Sec.\ \ref{sec:idk}.

\subsubsection{General procedure}

The ability of the correction procedure to overcome parameter inaccuracies, as suggested by the results in the perfect model scenario, inspired the comparison of three forecast models.
\begin{itemize}
	\item[(1)] tuned, uncorrected
	\item[(2)] untuned, corrected
	\item[(3)] tuned and corrected
\end{itemize}
Fig.\ \ref{fig:emcor} shows that empirical correction produces forecasts that remain useful longer, and demonstrate reduced error in this mock-operational setting, as well as in the perfect model scenario. Verification against both analysis and observations allows more confidence in the success of the procedure in aligning forecasts made by the low-dimensional model with the CFD simulated thermosyphon.

Additionally, the results constitute evidence that the correction procedure is more effective than fine-tuning of parameters for improving error statistics in this more realistic setting, as well as in the perfect model scenario. This is not to say that parameters should not be tuned, but rather that empirical correction is a cheaper and more effective avenue for reducing forecast uncertainty in the present context.

Empirical correction of the EM forecast model comes at the cost of decoupling qualitative dynamics from those of the true system, however. As shown in Fig.\ \ref{fig:rcdiffbyic}, a large region is created in which initial states have trajectories under the forecast model that behave completely different from how they would in the true system. An example of such a trajectory appears in Fig.\ \ref{fig:falsestab}. While the general empirical correction procedure reduces forecast error for many initial states, it produces a forecast model that is entirely useless for others. By introducing dynamical knowledge into the correction procedure, the results of which appear in the next section, we greatly reduce the number of such initial states, and also further reduce average forecast error.

\begin{figure}[htp]
     \centering
     \subfigure{
          \label{fig:acall}
          \hspace{.001in}
          \includegraphics[width=.88\columnwidth]{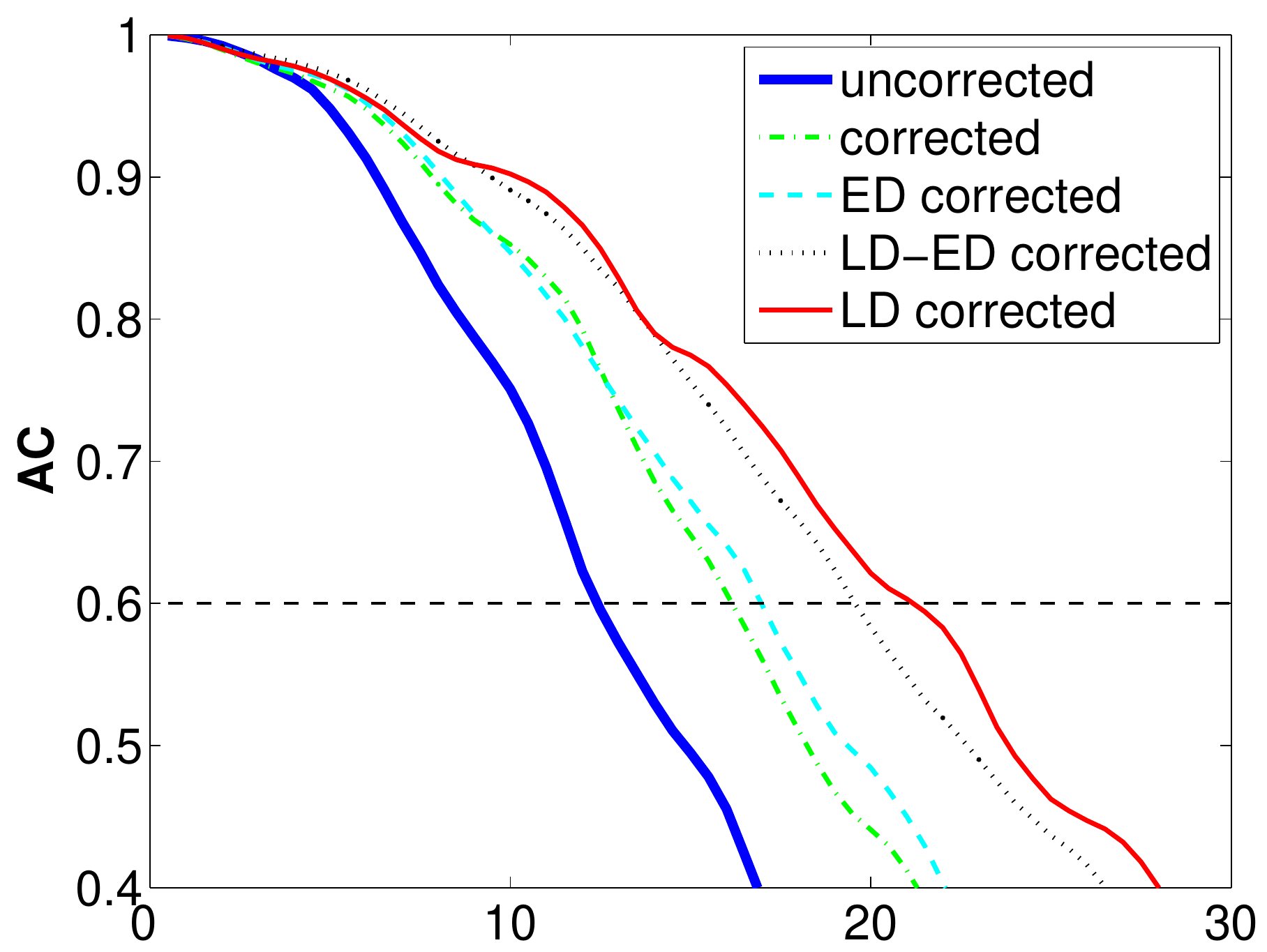}}  \\
     \vspace{-.2in}
     \subfigure{
          \label{fig:mfreall}
          \includegraphics[width=.9\columnwidth]{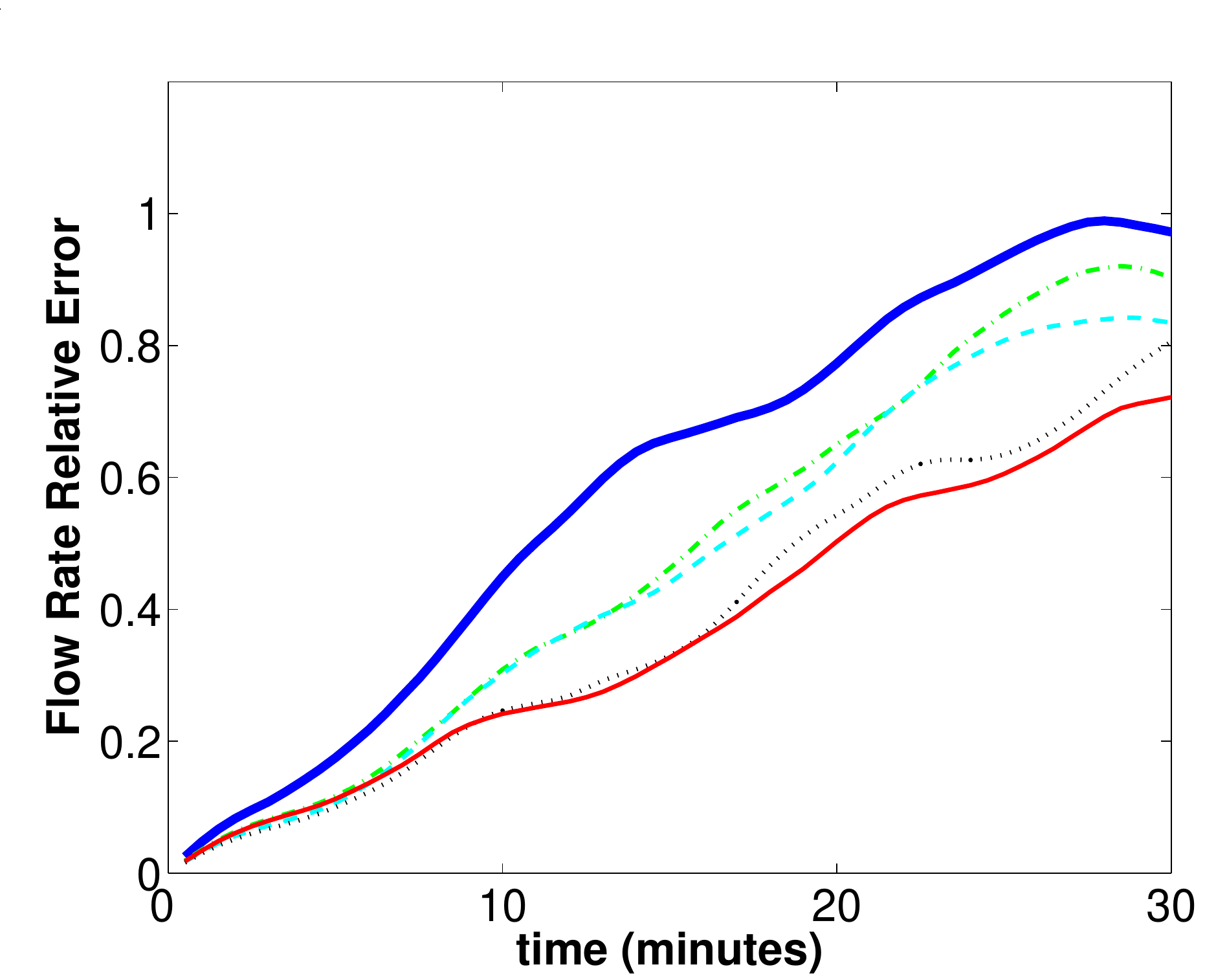}} \vspace{-.1in}
     \caption{\label{fig:acmfreall}  (Color online) Average AC (top) and mass flow-rate relative error (bottom) over 5000 trials for models that were: uncorrected (thick solid blue), corrected (green dash-dot), equilibrium-dependent (ED) corrected (cyan dash), lobe and equilibrium-dependent (LD-ED) corrected (black dot), and lobe-dependent (LD) corrected (solid red). Note that the models are listed in order of increasing performance, as measured by these error statistics. Lobe-dependent correction, i.e., correcting with bias terms and state-dependent operators that depend on the current flow regime direction, produces forecasts that are useful for almost twice as long as those made by the uncorrected model, on average. When compared to the dynamically uninformed correction scheme (green dash-dot), lobe-dependent correction more than doubles the improvement over the uncorrected model. We note that equilibrium-dependent correction does not produce much improvement in forecast error over the dynamically uninformed procedure, and hinders the performance when applied simultaneously with lobe-dependent correction.}  
\end{figure}

\subsubsection{Dynamically informed correction}

In an effort to produce forecast models that more accurately reflect the qualitative dynamics of the true system, two types of system-specific dynamical knowledge were directly incorporated into the empirical correction procedure: (1) flow regime direction of the current system-state, and (2) distance between the state and the nearest convective equilibrium solution. These dynamical cues were incorporated individually, and also simultaneously, resulting in three dynamically informed, empirically corrected forecast models, as outlined in Sec.\ \ref{sec:idk}. 

Fig.\ \ref{fig:acmfreall} shows average AC and mass flow-rate relative error over 5000 trials for the three dynamically informed and corrected models, as compared to the original biased model and the dynamically uninformed, corrected model. As described in the caption to the figure, encoding the current flow regime into the correction procedure results in a forecast model that is useful for almost twice as long as the original biased model, and doubles the improvement that was gained by dynamically uninformed emprical correction. Encoding the distance to the nearest equilibrium solution, on the other hand, does not greatly prolong usefulness, and in fact reduces it slightly when applied simultaneously with lobe-dependent correction.

\begin{widetext}

\begin{figure}[htp]
     \centering
     \includegraphics[width=.95\columnwidth]{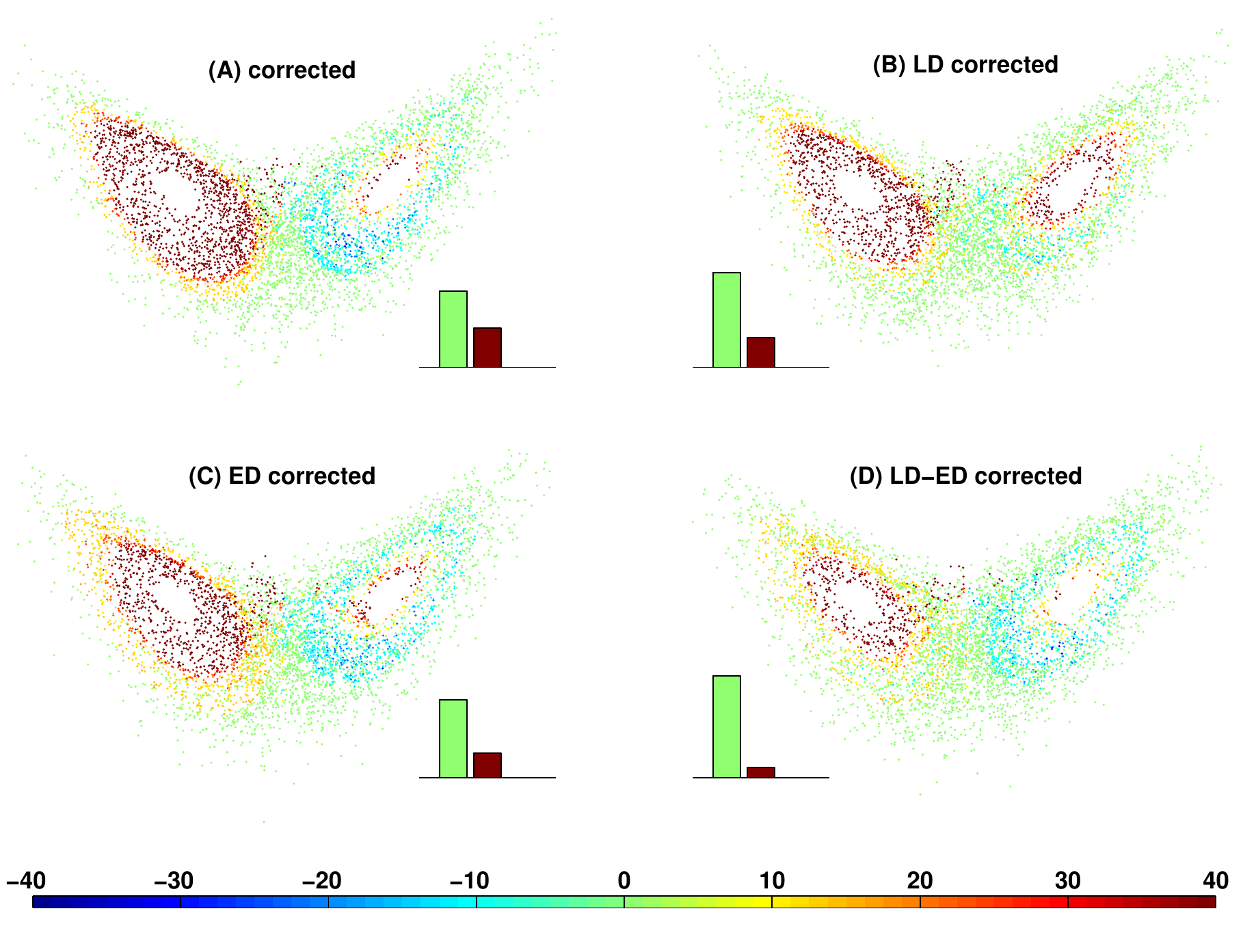} \vspace{-.2in}
     \caption{\label{fig:rcdiffbyicALL} (Color online) Difference (in minutes) between predicted and actual time of first flow reversal, plotted by initial state, for the (A) corrected, (B) lobe-dependent corrected, (C) equilibrium-dependent, and (D) lobe \underline{and} equilibrium-dependent models. As in Fig.\ \ref{fig:rcdiffbyic}, the difference was taken $t_{\text{model}} - t_{\text{actual}}$, so that positive values (towards red) indicate late predictions while negative values (towards blue) indicate early predictions. Inset partial histograms show the number of forecasts (out of 5000) predicting the first flow reversal within 3 minutes of the truth (green) and predicting that a flow reversal will never happen (red); the axes have the same scale for comparison, and the bar colors correspond to the dot colors.  (B) Applying lobe-dependent correction (as opposed to (A) dynamically uninformed correction) increases the number of initial states for which the first flow reversal is predicted accurately, visible as a larger number of green dots (bigger green bar). Also, although the region of dark red (initial states attracted to convective equilibrium) in the left lobe has decreased in size, the one in the right lobe has inflated, again as compared to the top left. (bottom left) Equilibrium-dependent correction shrinks the left-lobe red region without inflating the one in the right lobe. However, it maintains the region of initial states for which flow reversal predictions are slightly early (light blue), which was reduced by the lobe-dependent correction. (bottom right) Applying the simultaneously lobe and equilibrium dependent correction, in which there are four different correction regions, the forecast model demonstrates the smallest region of initial states whose qualitative dynamics are different from the CFD simulated thermosyphon. }  
\end{figure}

\end{widetext}

In Fig.\ \ref{fig:rcdiffbyicALL}, we see how the three forecast models resulting from dynamically informed empirical correction compare with the dynamically uninformed, corrected model, with regard to matching the qualitative behavior of the true system. We see a trend of improvement, characterized by smaller regions of states whose dynamics are different in the models than in the CFD simulated thermosyphon (dark red regions), as we apply first lobe-dependent, then equilibrium-dependent, and finally simultaneously lobe and equilibrium-dependent correction. 

Results shown in Figures \ref{fig:acmfreall} and \ref{fig:rcdiffbyicALL} suggest that encoding flow regime direction into the correction procedure primarily enhances forecast statistics, while encoding distance from equilibrium primarily enhances dynamical matching. Simultaneous inclusion of the two types of dynamical cues results in the best dynamical matching, with only a slight cost in average forecast accuracy. For further evidence of this summary conclusion, consider Table \ref{tab:1}.  

\begin{table}[]
\caption{\label{tab:1} Median absolute differences between predicted and actual times of first flow reversal (row 1), along with percentage of trials for which the first flow reversal was predicted within 1 and 2 minutes, (rows 2 and 3, respectively), for the uncorrected model (M), dynamically uninformed corrected model (CM), lobe-dependent corrected model (LD), equilibrium-dependent corrected model (ED), and simultaneously lobe and equilibrium-dependent corrected model (LD-ED).} 
\begin{tabular}{|c|c|c|c|c|c|} \hline &&&&& \vspace{-.1in}  \\
  & M & CM & LD & ED & LD-ED \\ \hline &&&&& \vspace{-.1in}  \\
  median & 4 & 10 & 1.5 & 10.5 & 1.5 \\ \hline &&&&& \vspace{-.1in}  \\
  \% within 1min. & 38.26 & 38.54 & 49.12 & 38.68 & 48.64 \\ \hline &&&&& \vspace{-.1in}  \\
  \% within 2min. & 47.04 & 43.40 & 54.38 & 43.58 & 57.08 \\ \hline 
\end{tabular}
\end{table}

In the first row, the median absolute differences between predicted and actual times of first flow reversal over the 5000 trials are listed for the uncorrected model (M), dynamically uninformed corrected model (CM), lobe-dependent corrected model (LD), equilibrium-dependent corrected model (ED) and simultaneously lobe and equilibrium-dependent model (LD-ED). The means are highly skewed for the CM, LD and ED models, due primarily to the large basins of attraction for the left-lobe equilibrium solution, see Fig.\ \ref{fig:rcdiffbyicALL} top left, right and bottom left, respectively, and thus we show only the medians. The medians show that both the LD and the LD-ED corrected models predict the first flow reversal more accurately than the uncorrected model. The bottom two rows show the percentage of the 5000 trials for which the first flow reversal was predicted within 1 and 2 minutes, respectively, for each of the models. Again the LD and LD-ED corrected models show the best performance. Note that the LD-ED model boosts both the 1 and 2-minute success rates by approximately 10\% over the uncorrected model. 

We note, however, that even the LD-ED model exhibits a spuriously stable convective equilibria in each lobe, Fig.\ \ref{fig:rcdiffbyicALL}. Incorporating dynamical knowledge of the true system in the correction procedure, at least through the partitioning of state-space as we have done here, is not enough to avoid the stabilizing effect of empirical correction on the equilibria of the EM model explained in Sec.\ \ref{sec:eqstab}. It is plausible that models of other systems with weakly repelling (attracting) equilibria might be subject to similar stabilizing (destabilizing) effects under empirical correction. Such effects might be mitigated by the strategy employed in the present work, but probably not avoided.

\section{\label{conc}Concluding Discussion}
It is apparent from the results of this work that the empirical correction technique tested here is successful in reducing forecast error in the low-dimensional setting. In both the perfect model scenario and the mock-operational experiment, improved error statistics and prolonged usefulness of forecasts were demonstrated by the corrected models. Furthermore, the empirical correction procedure was shown to provide greater improvement in average forecast accuracy than fine-tuning of parameters. However, this positive result comes at the cost of altering some important dynamical characteristics of the model. Details and implications of these results are discussed in the next two sections. In the third section, we make an observation about the relative impact of state-independent and state-dependent correction, in the present context, to conclude the paper.

\subsection{\label{ecvpt}Empirical correction vs.\ parameter tuning}

In the perfect model scenario, empirical correction of models with the greatest parameter error provided forecasts with smaller short-term average error, and that were useful for longer, than those made by the uncorrected models with least parameter error. Similarly, empirically correcting the EM model with 10\% error in \textsl{every} parameter (measured from tuned values) produced a forecast model that outperforms the tuned, but uncorrected model in predicting the CFD simulated thermosyphon. In each experiment, superior performance is evident in terms of anomaly correlation and average forecast error, measured against 3DVar analysis, and is verified against observed mass flow-rate of the CFD simulation in the toy climate experiment.

We do not present these results as evidence that empirical correction should replace parameter tuning, nor even that the former is better than the latter in any well-defined sense. However, the results do suggest that empirical correction could be a viable complement to the tuning of model parameters. Particularly as degrees of freedom become large, for example $N \approx 10^{10}$ in some currently operational numerical weather models, the computational cost of parameter tuning is very large in comparison to that of empirical correction, when appropriately modified for such models (see \cite{danforth2007,danforth2008singularvals}). An example of such a modification is to compute the first $m \ll N$ principal components (PCs) of $\mathbf{L}$ by singular value decomposition (SVD), where $m$ is determined so that a certain percentage of the state covariance is explained by these first $m$ PCs. A combined tuning/correction approach could reduce the number of model integrations necessary in the parameter tuning process without sacrificing model accuracy. However, the dynamical ramifications of this strategy must be considered, as we explain in the next section.  

\subsection{\label{esvsd}Error statistics vs.\ system dynamics}

The reduction in average forecast error provided by empirical correction belies fundamental dynamical disturbances born out of the correction procedure. Stabilization of equilibrium solutions, which results in large regions of initial states for which the corrected model forecasts behavior in opposition to that of the true system, follows from the dynamical modification imposed by empirical correction. In addition, the symmetry of the model system is broken by empirical correction. Though these costs can be mitigated somewhat by hard-wiring system-specific dynamical cues into the correction procedure, they can not be eradicated without more fundamental alterations of the technique, e.g., forcing the bias term $\mathbf{b}$ and Leith operator $\mathbf{L}$ to preserve system symmetry. In operational practice, empirical correction is known to introduce imbalances, e.g. violating geostrophy, necessitating some mechanism for smoothing the flow into a physically viable region of state space.

In fact, it may be impossible to avoid all dynamical inaccuracies resulting from empirical correction, and even if theoretically possible, it would likely be impractical to do so in any operational setting. In considering the application of the technique in operational settings, then, it must be determined if the effects of misrepresented dynamics can be reduced to a tolerable level on a case-by-case basis. In the ideal situation, regions of state-space that would be dynamically misrepresented under empirical correction could be reduced, by minor modifications to the correction procedure, to encompass only unrealistic or unlikely physical states. In any case, the technique presented in this study should not be applied without such considerations. 

\subsection{\label{bcvlo}Bias correction vs.\ Leith operator}

State-independent error correction by itself produces almost no improvement in any of the forecast models in this study (not shown). This is in contrast to what has been observed in operational weather and climate model studies, where state-independent bias correction typically outperforms state-dependent correction in reduction of forecast errors \cite{yang2008}. The inaccuracies of ad-hoc forcings included in such models to compensate for external and/or irresolvable phenomena (e.g. solar and cloud forcings, respectively) are likely responsible for a large component of the bias. In light of the lack, or at least minimal nature of such external and sub-gridscale influences in the toy models considered here, the ineffectiveness of bias correction is logically consistent with this explanation. 

The state-dependent Leith operator is entirely responsible for the success of the corrected models in this study. In the perfect model scenario this makes sense because the difference between the forecast models and the ``truth'' model are inherently multiplicative, i.e. the parameters are coefficients weighting the interaction between state-variable values and thus resulting errors \textsl{must} depend on state. For the EM model of the CFD system, it seems that errors resulting from the low dimensionality of the forecast model may also be multiplicative in nature. If this is the case, state-dependent correction may reduce error patterns in operational models that result from reduced dimensionality, e.g. coarse resolution. The correction will not likely compensate for processes that are irresolvable due to coarse resolution, but rather may reduce the propagation of error resulting from the omission of such phenomena. This hypothesis is consistent with demonstrated improvement of local behavior in state-dependent corrected atmospheric models with $N \approx 10^{5}$ degrees of freedom \cite{danforth2007}.

In previous studies of state-dependent correction in models that are much more realistic than those considered here, resulting error reduction has been minuscule in comparison to what is achieved by bias correction. However, this is not cause to reject the usefulness of parameterizing state-dependent error. Though globally averaged error reduction may not be significant, improvement in the local behavior of models can have a large impact on forecast uncertainty, particularly in an ensemble strategy where state-dependent correction can increase the spread in previously unsampled state-space directions. 

\section*{Acknowledgements}

Simulations and experiments were performed on the Vermont Advanced Computing Center (VACC) 1400-processor cluster, an IBM e1350 High Performance Computing system. The authors would like to thank Ross Lieb-Lappen for assistance in preparing this manuscript. This study was supported by NSF grant \#DMS0940271, the Mathematics and Climate Research Network, and a National Aeronautics and Space Administration (NASA) EPSCoR grant.


\end{document}